\documentclass{article} 
\usepackage{iclr2025_conference,times}

\usepackage{amsmath,amsfonts,bm}

\def\eqref#1{equation~\ref{#1}}

\def\1{\bm{1}}

\DeclareMathAlphabet{\mathsfit}{\encodingdefault}{\sfdefault}{m}{sl}
\SetMathAlphabet{\mathsfit}{bold}{\encodingdefault}{\sfdefault}{bx}{n}

\usepackage{hyperref}
\usepackage{url}
\usepackage{multirow}
\usepackage{booktabs} 
\usepackage[table]{xcolor}
\usepackage{graphicx}
\usepackage{caption}
\usepackage{etoc}
\usepackage{xurl}
\hypersetup{hidelinks}
\usepackage{xcolor}
\usepackage[most]{tcolorbox}
\etocdepthtag.toc{mtchapter}
\etocsettagdepth{mtchapter}{subsubsection}
\etocsettagdepth{mtappendix}{none}

\newcommand{\modelicon}[1]{%
  \makebox[1.35em][c]{%
    \raisebox{-0.15em}{%
      \includegraphics[width=1.05em,height=1.05em,keepaspectratio]{fig/icon/#1}%
    }%
  }\hspace{0.25em}%
}
\newcommand{\linkicon}[1]{%
  \raisebox{-0.2em}{\includegraphics[height=1.25em,keepaspectratio]{fig/icon/#1.png}}\hspace{0.25em}%
}
\newlength\savewidth
\newcommand{\tablestyle}[2]{\setlength{\tabcolsep}{#1}\renewcommand{\arraystretch}{#2}\centering\footnotesize}
\renewcommand{\paragraph}[1]{\vspace{1.25mm}\noindent\textbf{#1}}
\newcommand\blfootnote[1]{%
  \begingroup
  \renewcommand\thefootnote{}%
  \NoHyper\footnotetext{\hspace*{-1.8em}\ignorespaces#1}\endNoHyper%
  \endgroup
}

\title{Active-SWE: Benchmarking Coding Agents for Proactive Bug Fixing without Issue Reports}

\newcommand{\activemark}[1]{\raisebox{0.7ex}{\fontsize{7}{8}\selectfont\mdseries #1}}
\newcommand{\activeauthor}[2]{\textbf{#1}\activemark{#2}}
\newcommand{\makeactivefront}{%
\noindent\begin{minipage}{\textwidth}
  \centering
  \raisebox{-0.2\height}{\includegraphics[height=4em]{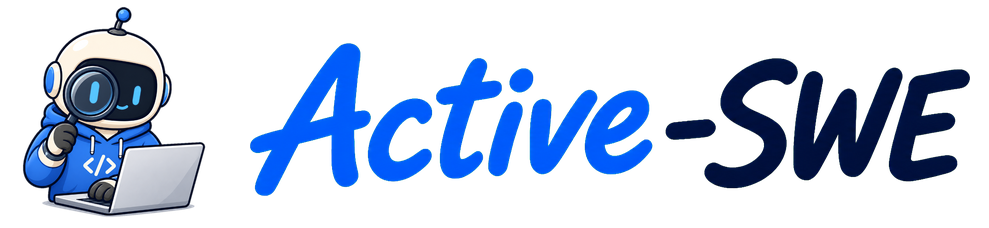}}\par
  \vspace{2.5mm}
  {\fontsize{16}{19}\selectfont\bfseries
    Benchmarking Coding Agents for Proactive Bug Fixing without Issue Reports\par
  }
  \vspace{3.5mm}
  
  {\fontsize{11}{15}\selectfont
    \activeauthor{Haobin Li}{1,*}\quad
    \activeauthor{Ping Deng}{2,*}\quad
    \activeauthor{Weizhong Qian}{2}\quad
    \activeauthor{Liang Jiang}{3}\par
    \vspace{1mm}
    \activeauthor{Zhenyu Huang}{1,\textdagger}\quad
    \activeauthor{Mouxing Yang}{1,\textdagger}\quad
    \activeauthor{Xi Peng}{1,\textdagger}\par
  }
  
  \vspace{2mm}
  {\fontsize{10}{14}\selectfont
    \activemark{1}Sichuan University\hspace{1.2em}
    \activemark{2}University of Electronic Science and Technology of China \\
    \activemark{3}Independent Researchers\par
  }
  
  \vspace{4mm}
  
  \begin{minipage}{\linewidth}
  \linespread{1.1}\fontsize{10}{12}\selectfont
  \setlength{\parindent}{0pt}
  Coding agents powered by large language models (LLMs) are increasingly adopted in software engineering (SWE) scenarios, capable of fixing a specific bug in large-scale codebase.
  However, existing SWE benchmarks typically assume that high-quality issue reports with detailed information are always available, which is easily violated in practice due to the complexity of report acquisition and curation.
  To address this, we introduce \textbf{Active-SWE}, a benchmark for evaluating coding agents on proactively discovering and fixing multiple bugs without report guidance, covering 1,663 tasks across six bug categories and eight languages.
  Beyond shifting the focus from existing reactive bug fixing to proactive bug fixing, Active-SWE enables a more in-depth evaluation by expanding the scope from fixing a specific recorded bug to multiple-bug fixing and potential bug discovery scenarios.
  To construct Active-SWE, we propose a novel difficulty-aware task formulation pipeline with a dual-track evaluation framework, facilitating comprehensive evaluation of proactive bug-fixing capability.
  Extensive experiments reveal that most state-of-the-art coding agents struggle with proactive bug-fixing tasks, demonstrating limited performance in locating and resolving recorded bugs, handling multiple bug fixing scenarios, and discovering valid potential bugs.
  \end{minipage}
  
  \vspace{3mm}
  
  {\normalfont
    \linkicon{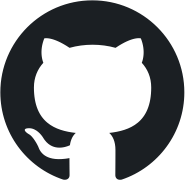}\href{https://github.com/XLearning-SCU/Active-SWE}{\textbf{GitHub}}
    \qquad\qquad
    \linkicon{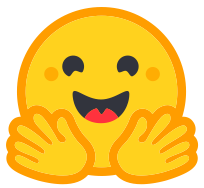}\href{https://huggingface.co/datasets/XLearning-SCU/Active-SWE}{\textbf{Benchmark}}
    \qquad\qquad
    \linkicon{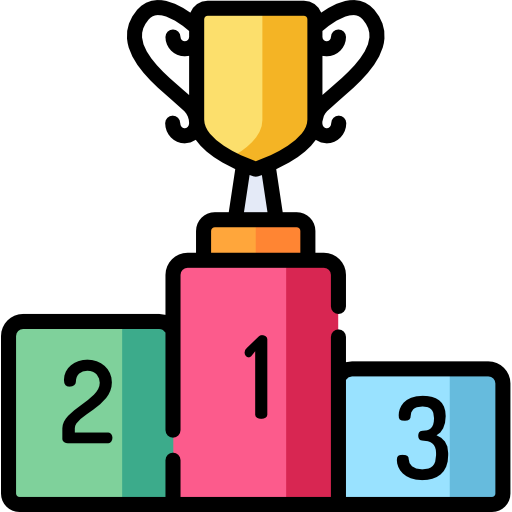}\href{https://hbinli.github.io/Active-SWE/}{\textbf{Leaderboard}}
  }
\end{minipage}%
}

\iclrfinalcopy 

\begin{document}

\thispagestyle{plain}
\vspace*{-2.5em}
\makeactivefront
\vspace{4mm}

\blfootnote{%
  \footnotesize
  \activemark{*}Equal contribution. \quad
  \activemark{\textdagger}Corresponding authors.\\[0.2em]
  Emails: \{haobinli.gm, pingdeng001, zyhuang.gm, yangmouxing, pengx.gm\}@gmail.com
}

\section{Introduction}

\begin{figure*}[t]
    \centering
    \includegraphics[width=0.98\linewidth]{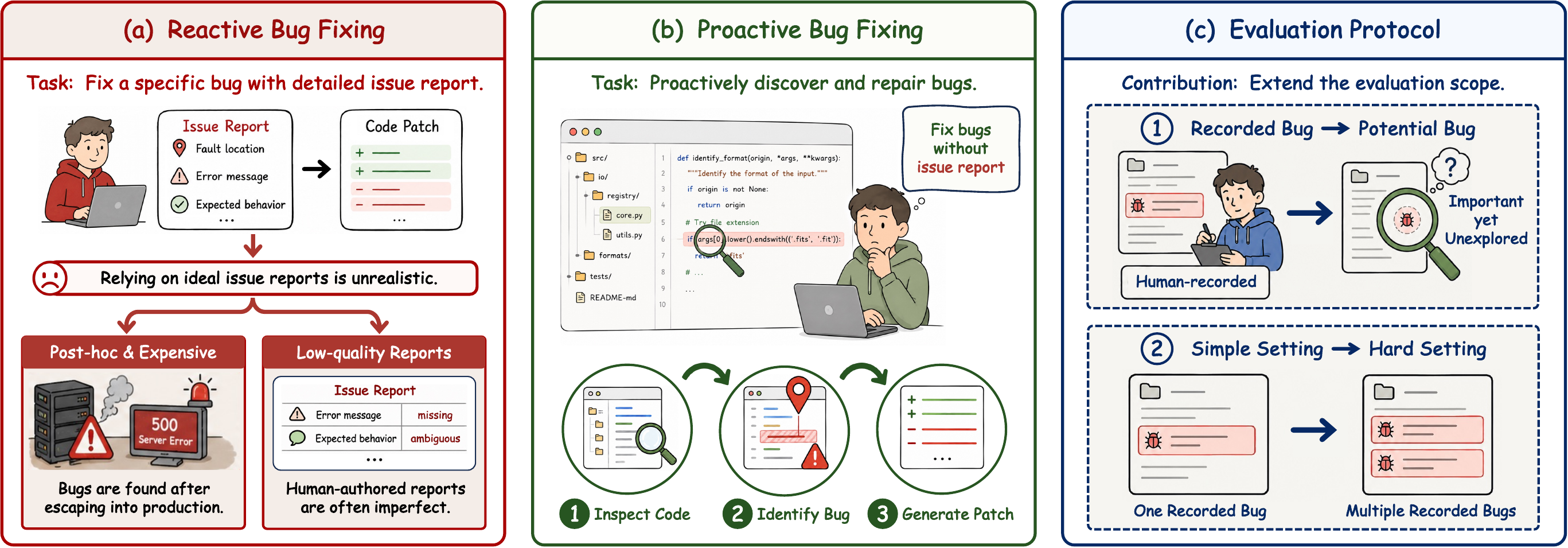}
    \caption{
    (a) \textbf{Reactive Bug Fixing:} 
    existing benchmarks aim to fix a specific bug with detailed issue reports, largely overlooking that such desirable reports are often unavailable due to costly bug identification and unreliable issue descriptions.
    (b) \textbf{Proactive Bug Fixing:} we introduce Active-SWE to evaluate coding agents on proactive bug discovery and repair without issue reports, posing greater demands on code inspection, bug identification, and repair capabilities.
    (c) \textbf{Comprehensive Evaluation:} two novel evaluation dimensions are designed for the comprehensive assessment of proactive bug-fixing capability:
    i) validating both recorded bug fixing and potential bug discovery;
    ii) supporting bug-fixing tasks that involve multiple bugs rather than a specific fix.
    }
    \label{fig: fig1}
\end{figure*}

Coding agents~\citep{swe-agent,codeagent2} driven by Large Language Models (LLMs)~\citep{llm1,llm2} have been increasingly applied to Software Engineering (SWE) scenarios~\citep{swe1,swe2}, with the research community expanding from simple programming tasks to real-world software engineering tasks such as fixing a specific bug in large codebase.
To comprehensively evaluate coding agents in repository-level SWE scenarios, representative benchmarks construct instances from GitHub issues and their corresponding pull requests (PRs), \textit{e.g.}, SWE-bench Verified~\citep{swe-bench} and SWE-bench Pro~\citep{swe-bench-pro}, where coding agents are typically tasked with generating a code patch for the specific bug described in the human-written issue report.

However, existing SWE benchmarks~\citep{swe-bench-live,multi-swe-bench} evaluate coding agents under the ideal assumption that human-authored issue reports are available and contain desirable debugging information, \textit{e.g.}, fault locations, error messages, and expected behavior, which is daunting to satisfy as illustrated in Fig.~\ref{fig: fig1}(a).
On the one hand, the acquisition of issue reports is inherently post-hoc and prohibitively expensive, \textit{i.e.}, bugs are typically identified after escaping into production, leading to severe consequences such as resource waste, service outages, and security vulnerabilities. 
For instance, the 2024 CrowdStrike outage~\citep{onehand_case1} was caused by an out-of-bounds read bug in the Content Interpreter, crashing millions of Windows devices and incurring billions of dollars in losses~\citep{onehand_case2,onehand_case3}.
On the other hand, due to the lack of expert knowledge and standardized practices, human-authored descriptions in issue reports are often ambiguous and incomplete, with insufficient information to resolve potential problems.
In particular, recent analyses~\citep{otherhand_case1,otherhand_case2} from OpenAI reveal that even carefully curated benchmarks such as SWE-bench Verified and SWE-Bench Pro contain non-negligible low-quality issue reports, resulting in tasks that are infeasible to resolve.
Taken together, existing benchmarks focus on evaluating the reactive bug-fixing capability of coding agents, which relies on human involvement and high-quality issue reports.

In light of the limitation, we pose a more general and challenging research question for SWE scenarios: 
\begin{center}
\begin{minipage}{0.75\linewidth}
\itshape
Can coding agents proactively discover and resolve bugs before they are reported or even hinted at by developers?
\end{minipage}
\end{center}
 
As shown in Fig.~\ref{fig: fig1}(b), this capability involves inspecting the code repository, identifying potentially multiple bugs, and generating corresponding fixes without issue reports, which we refer to as \textit{proactive bug fixing}. 
To answer this question, we introduce Active-SWE, which might be the first benchmark to comprehensively evaluate the proactive bug-fixing capability of coding agents.
Different from existing benchmarks, Active-SWE sheds light on the capacity of coding agents to act as active bug fixers, thus facilitating a more in-depth evaluation from two perspectives:
i) \textbf{multiple bug scenarios}: existing benchmarks focus on resolving one specific bug, while Active-SWE further validates the capability to discover and resolve multiple bugs in proactive bug fixing;
ii) \textbf{potential bug scenarios}: beyond the evaluations on recorded bugs, Active-SWE extends the evaluation scope to reveal important yet under-explored bugs in real-world repositories.

In this paper, we introduce a novel pipeline that systematically constructs Active-SWE from real-world SWE issues and evaluates it in a rigorous and reproducible manner.
First, we adopt a taxonomy-driven data curation module to derive executable repository snapshots with discoverable bugs from GitHub PRs, which supports reliable benchmark construction and evaluation.
Second, we propose a novel difficulty-aware task formulation module that reformulates raw PRs as proactive bug-fixing tasks and then derives tasks involving multiple bug fixes.
Third, we design a dual-track evaluation framework to comprehensively evaluate the bug-fixing capability of coding agents from both the recorded bug repair and potential bug discovery perspectives.
In summary, the major contributions and novelties of this work are given as follows.
\begin{itemize}
    \item We propose Active-SWE, which could be one of the first benchmarks that shifts the focus from reactive to proactive bug fixing. Different from existing benchmarks, Active-SWE evaluates the capability of coding agents to proactively discover and fix bugs without issue reports, covering 1,663 tasks across diverse bug categories and programming languages.
    \item For comprehensive evaluations of proactive bug-fixing capability, we propose a novel benchmark construction pipeline with a dual-track evaluation framework, extending evaluation scope to both multiple bug fixing and potential bug discovery scenarios.
    \item Extensive experiments on the Active-SWE demonstrate that even state-of-the-art coding agents still struggle with proactive bug-fixing tasks, especially in locating and resolving recorded bugs, handling hard tasks with multiple bugs, and discovering valid potential bugs.
\end{itemize}

\section{Related Work}
\subsection{Benchmarks for Software Engineering}
Real-world software engineering requires nuanced reasoning over large repositories and performing complex code modifications under diverse tasks~\citep{swe-bench-multimodal,swe-atlas}.
According to the characteristics of tasks, existing SWE benchmarks could be broadly grouped into three categories:
i) issue-solving benchmarks~\citep{swe-bench-live}, which focus on evaluating whether LLMs could resolve bugs or other software problems described in the issue;
ii) test-generation benchmarks~\citep{testexplora,swt-bench}, where LLMs are tasked to derive executable tests that capture expected behaviors from issue descriptions or curated documents;
iii) feature-development benchmarks~\citep{featurebench,swe-atlas}, which require LLMs to translate high-level development documents into function-level or repository-level implementations.

In this work, we systematically evaluate a highly-practical yet less-explored task, \textit{i.e.}, proactive bug fixing, which involves discovering and repairing bugs in the repository without external guidance.
In other words, existing SWE benchmarks typically rely on carefully-curated issues or documents that would specify predefined objectives and provide detailed information.
In contrast, the bug-fixing task aims to first identify fixing objectives and then implement them without human involvement, which poses higher demands on code exploration, objective identification, and code implementation capabilities of coding agents.

\subsection{Coding Agents}
Over the last few years, LLM-based coding agents~\citep{codeagent1,codeagent2,codeagent3} have emerged as the dominant paradigm in real-world software engineering, where LLMs are equipped with tools to interact with the development environments over multiple turns.
In particular, recent advances on coding agents have demonstrated substantial performance gains over tool-free LLMs across various challenging tasks, including repository-level code repair and generation.
With the rapid development of agent harnesses~\citep{swe-agent,openhands}, coding agents are able to leverage command-line tools to search and inspect repositories, edit source code, and execute tests, thus supporting various complex SWE tasks~\citep{codeagent_swe1,codeagent_swe2}.
In this work, we adopt diverse instructions and inputs to endow coding agents with the ability to construct the benchmark and evaluate the proactive bug-fixing capability.


\section{Active-SWE}
Active-SWE is a benchmark for evaluating the proactive bug-fixing capability of coding agents, which comprises 1,663 tasks spanning six major bug categories and eight programming languages.
In this section, we will introduce the task formulation, data curation pipeline and the evaluation framework of Active-SWE.

\subsection{Task Formulation}

Let $r_i$ denote the repository snapshot of the $i$-th bug fixing instance, $\mathcal{F}_i$ indicate the corresponding set of files to be inspected, and $\hat{\mathbf{b}}_i$ represent the recorded bugs contained in $\mathcal{F}_i$.
For a given repository snapshot $r_i$, the goal of the bug-fixing task is to generate a code patch $c_i$ for resolving the discovered bugs within the review scope $\mathcal{F}_i$, \textit{i.e.},
\begin{equation}
    c_i = \mathcal{L} \left(r_i, \mathcal{F}_i, \mathcal{T}_{\text{R}} \right),
    \label{eq: task_generate}
\end{equation}
where $\mathcal{T}_{\text{R}}$ denotes the unified bug-fixing task template and $\mathcal{L}$ denotes the evaluated LLM. 
After that, we adopt the following test-driven evaluation protocol to verify the effectiveness of the predicted patch $c_i$, \textit{i.e.},
\begin{equation}
    \mathcal{P}(c_i \oplus r_i,\mathbf{t}_i)\in[0,1],
    \label{eq: task_test}
\end{equation}
where $\mathbf{t}_i$ indicates the test set, $\oplus$ denotes the patch application operation, $\mathcal{P}(\cdot)\in [0,1]$ denotes the test pass rate, and $\mathcal{P}(c_i \oplus r_i,\mathbf{t}_i)=1$ \textit{i.f.f.} all tests in $\mathbf{t}_i$ are passed successfully after applying $c_i$ to $r_i$.
Different from the existing bug-fixing benchmarks that rely on the costly, post-hoc issue reports, the bug-fixing task shifts the focus towards proactively discovering and repairing bugs with a unified bug fixing template, which is more general and practical in real-world SWE scenarios.
In the following, we will elaborate on the construction pipeline of the Active-SWE and the evaluation framework for validating the bug fixing capabilities of coding agents.

\begin{figure*}[t]
    \centering
    \includegraphics[width=0.98\linewidth]{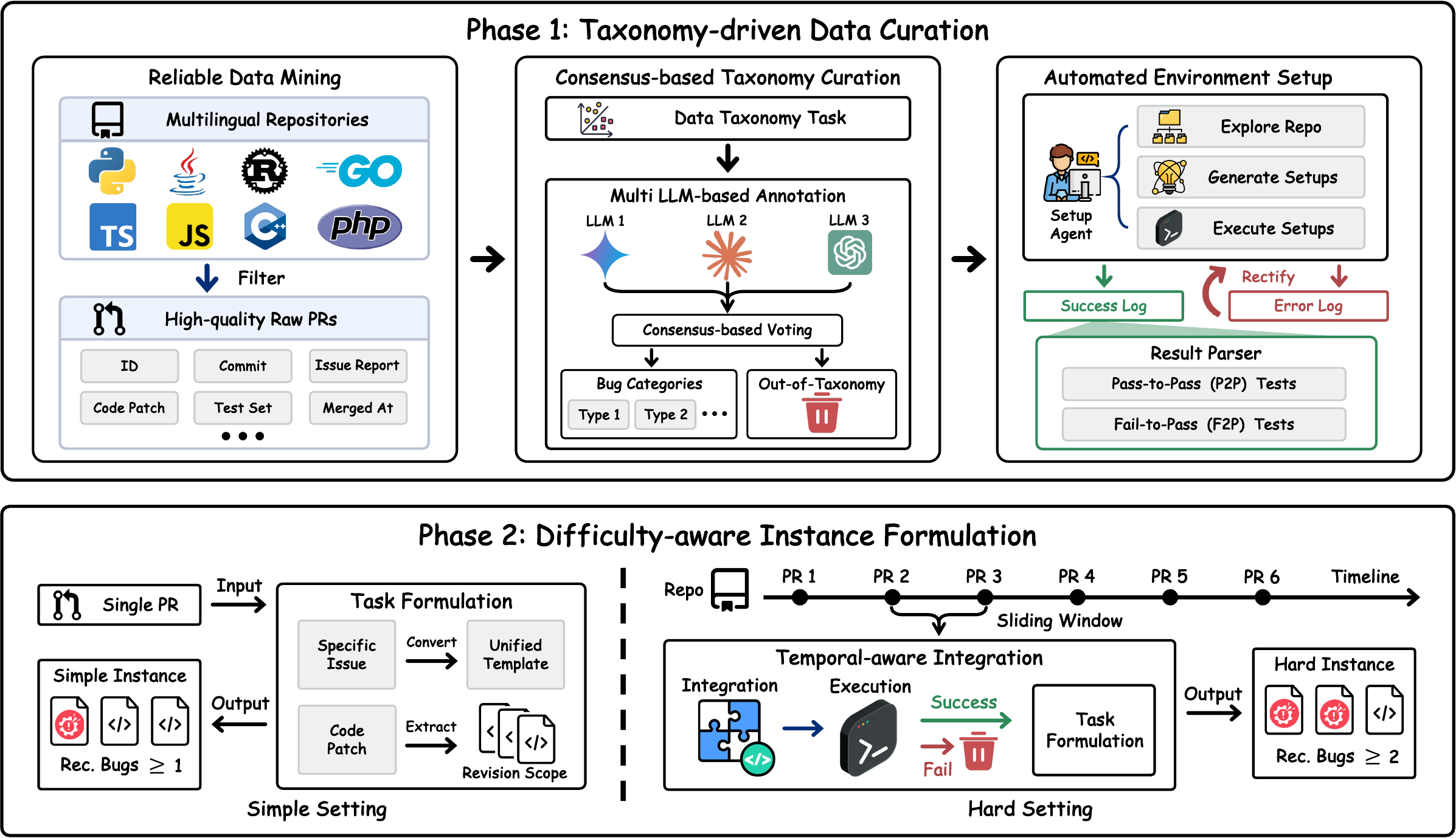}
    \caption{
    \textbf{Construction pipeline of Active-SWE.} 
    The pipeline consists of two phases:
    i) we crawl high-quality multilingual PRs from GitHub, identify discoverable bug-fixing PRs, and build executable environments with a setup agent;
    ii) we reformulate mined PRs into proactive bug-fixing tasks, \textit{i.e.}, each simple instance is derived from a single PR, while each hard instance is constructed by integrating temporally adjacent PRs with a sliding window mechanism.
    }
    \label{fig: fig2}
\end{figure*}

\subsection{Taxonomy-driven Data Curation}
To collect data for constructing Active-SWE, it is essential to curate executable repository snapshots with realistic and discoverable bugs.
Accordingly, we propose a taxonomy-driven data curation pipeline consisting of reliable data mining, consensus-based taxonomy curation, and automated environment setup, as illustrated in Fig.~\ref{fig: fig2}.

\textbf{Reliable Data Mining.}
GitHub provides a rich source of real-world bug fixing histories, where PRs record how developers identify and repair bugs in evolving codebases. 
Since the raw PRs are usually noisy and poorly documented, we first mine high-quality PRs from open-source repositories.
Specifically, we select PRs from \textbf{87} popular open-source repositories across \textbf{eight} programming languages, each with at least \textbf{1,000} GitHub stars.
For each repository, we crawl its historical PRs and select those linked to GitHub issues with reports, reference code patch, reference test set, merged time, and so on.

\textbf{Consensus-based Taxonomy Curation.}
Although the mined PRs are of high quality, not all of them correspond to bug-fixing activities, \textit{e.g.}, some involve feature implementations or API extensions.
To remedy this, we propose a consensus-based curation mechanism for filtering out PRs without discoverable bugs.
Specifically, for a given PR with issue report $s_i$, reference code patch $\hat{c}_i$, and reference test set $\hat{\mathbf{t}}_i$, we employ multiple LLM-based annotators to assign taxonomy labels as follows,
\begin{equation}
    y_i^{k} = \mathcal{A}^k\left(s_i, \hat{c}_i, \hat{\mathbf{t}}_i, \mathcal{T}_{\text{C}}\right),
\end{equation}
where $y_i^{k}$ denotes the taxonomy label predicted by the $k$-th annotator $\mathcal{A}^{k}$, $\mathcal{T}_{\text{C}}$ denotes the taxonomy task template.
After that, we derive the final taxonomy label through consensus-based voting strategy, \textit{i.e.},
\begin{equation}
    y_i = \mathcal{V}\left(y_i^{1}, \ldots, y_i^{K}\right),
\end{equation}
where $K$ denotes the number of annotators and $\mathcal{V}(\cdot)$ denotes the voting operation. 
In the implementation, the taxonomy is defined with software engineering experts, consisting of \textbf{6} major bug categories and an additional \textit{Out-of-Taxonomy} category. 
PRs labeled as \textit{Out-of-Taxonomy} are discarded, while the remaining PRs are retained as valid ones.

\textbf{Automated Environment Setup.}
To facilitate faithful execution and evaluation of bug-fixing tasks, we employ an LLM-based setup agent to establish docker-based environments and test scripts for retained PRs.
Specifically, the setup agent adopts the following ReAct-style~\citep{react} process:
i) \textbf{explore} $r_i$ to identify build commands and test commands, then generate the dockerfile and test script; 
ii) \textbf{execute} the generated setups by building the docker image and running test script; 
iii) \textbf{rectify} dockerfile and test script based on execution feedback until success or the turn limit is reached. 
To facilitate the subsequent benchmark construction and evaluation, the setup agent is required to generate the test script that reports test-level outcomes.
According to the test-level outcomes, the reference test set $\hat{\mathbf{t}}_i$ could be decomposed into fail-to-pass tests $\hat{\mathbf{f}}_i$ and pass-to-pass tests $\hat{\mathbf{p}}_i$.

\subsection{Difficulty-aware Task Formulation}
\label{sec: difficulty_aware_task_formulation}
In this section, we elaborate on how to reformulate raw PRs into proactive bug-fixing tasks and construct tasks with different difficulty levels.

\textbf{Simple Setting.}
For the simple setting, each bug fixing instance is derived from a single valid PR, whose files pending review contain at least one recorded bug, \textit{i.e.}, $|\hat{\mathbf{b}}_i|\geq 1$.
To be specific, for a given PR, we discard issue report $s_i$ and instead treat a unified task template $\mathcal{T}_{\text{R}}$ as instruction, which prompts coding agents to proactively fix bugs.
After that, we extract the files pending review $\mathcal{F}_i$ from reference code patch $\hat{c}_i$, thus guaranteeing the review scope covers the recorded bug,
\begin{equation}
    \mathcal{F}_i = g\left(\hat{c}_i\right),
    \label{eq: extract_files}
\end{equation}
where $g(\cdot)$ denotes the file extraction operation.
The remaining artifacts are directly inherited from the PR to construct the simple instance $(r_i, \hat{\mathbf{t}}_i, \mathcal{F}_i, \mathcal{T}_{\text{R}})$.

\textbf{Hard Setting.}
For the hard setting, we adopt the temporal-aware integration mechanism to formulate tasks with multiple recorded bugs, where the review scope involves at least $M$ bugs, \textit{i.e.}, $|\hat{\mathbf{b}}_i|\geq M$.
The key insight is that an earlier repository snapshot may contain multiple bugs later addressed by subsequent PRs, enabling us to aggregate them for deriving hard instance.
Specifically, for each repository, we sort valid PRs by merge time and slide a temporal window over the ordered sequence, where $\mathbf{w}_i$ denotes the indices of PRs in the $i$-th window.
After that, we derive the integrated artifacts of the corresponding hard instance as follows,
\begin{equation}
    \hat{c}_i = \bigcup_{j\in\mathbf{w}_i}\hat{c}_j,
    \quad
    \hat{\mathbf{t}}_i = \bigcup_{j\in\mathbf{w}_i}\hat{\mathbf{t}}_j,
\end{equation}
where $\bigcup$ denotes the integration operation.
To verify the integrated artifacts $\hat{c}_i$ and $\hat{\mathbf{t}}_i$ are valid, we preserve the high-quality hard instance only if the following condition is satisfied, \textit{i.e.}, $\hat{\mathbf{f}}_i = \bigcup_{j\in\mathbf{w}_i}\hat{\mathbf{f}}_j$. 
In other words, the fail-to-pass tests of the hard instance are expected to be the union of those from the individual PRs.
Then, we extract the integrated review scope $\mathcal{F}_i$ according to Eq.~\ref{eq: extract_files}, while the other artifacts are inherited from the earliest PR in $\mathbf{w}_i$.

\subsection{Dual-Track Evaluation Framework}
As discussed in Introduction, the key to evaluating the proactive bug fixing capability of coding agents lies in two dimensions:
i) whether agents could discover and repair the recorded bugs;
ii) whether agents could reveal valid yet under-explored bugs.
To this end, we propose a novel dual-track framework to facilitate the comprehensive evaluation of coding agents.

\subsubsection{Recorded Bug Evaluation}
Existing issue-driven benchmarks rely on detailed debugging cues provided by issue reports, which significantly simplify the bug localization and repair process.
Since reports are unavailable in proactive bug-fixing tasks, it is necessary to diagnose the limitations of coding agents in terms of the two aspects.
To this end, we propose decoupled metrics for evaluating recorded bug fixing capabilities of coding agents, which treat the reference code patch $\hat{c}_i$ and the reference test set $\hat{\mathbf{t}}_i$ as the oracle for bug localization and code repair, respectively.

\textbf{Localization Recall (LR) and Precision (LP):}
we quantify localization capability with two fine-grained edit-matching metrics, \textit{i.e.}, 
\begin{equation}
    \mathrm{LR} =
    \frac{|\hat{\mathcal{H}}_i^{\mathrm{hit}}|}{|\mathcal{H}(\hat{c}_i)|},
    \quad
    \mathrm{LP} =
    \frac{|\mathcal{H}_i^{\mathrm{hit}}|}{|\mathcal{H}(c_i)|},
\end{equation}
where $c_i$ denotes the generated code patch in Eq.~\ref{eq: task_generate}, $\mathcal{H}(\cdot)$ extracts the edit hunks from the patch, and $\hat{\mathcal{H}}_i^{\mathrm{hit}}=\{\hat{h}_j\mid \hat{h}_j\in\mathcal{H}(\hat{c}_i),\exists h_k\in\mathcal{H}(c_i),\hat{h}_j\sim h_k\}$ and $\mathcal{H}_i^{\mathrm{hit}}=\{h_k\mid h_k\in\mathcal{H}(c_i),\exists \hat{h}_j\in\mathcal{H}(\hat{c}_i),h_k\sim\hat{h}_j\}$ with $\sim$ indicating that two hunks are matched, 
Such behavior not only treats the reference code patch as oracle for reliable localization capability estimation, but also decomposes the code patch into fine-grained units and thus enables scalable analysis in multi-bug scenarios.

\textbf{Resolved:} we employ the test-driven evaluation protocol for estimating the bug repair capability, \textit{i.e.},
\begin{equation}
    \mathrm{Resolved}
    =
    \mathbb{I}\left[
    \mathcal{P}(c_i\oplus r_i,\hat{\mathbf{t}}_i)=1
    \right],
\end{equation}
where $\mathbb{I}[\cdot]$ is an indicator function evaluating to $1$ \textit{i.f.f.} the condition is satisfied, and $\mathrm{Resolved}$ indicates whether the recorded bugs are fully addressed.

\subsubsection{Potential Bug Evaluation}
Beyond recorded bugs, Active-SWE further evaluates whether the potential bugs revealed in the generated code patch are valid.
However, evaluating potential bugs is quite challenging as no reference oracle is available.
Instead of exhaustingly collecting human-curated oracles, we reformulate the patch validation as the test generation problem, where revealed bugs are credible only if reproduced by valid test evidence.
To this end, we propose a trustworthy test-driven evaluation mechanism, which tasks agents with generating tests for reproducing the revealed bugs.
Specifically, given a predicted code patch $c_i$, the evaluated LLM is instructed to generate a set of tests $\mathbf{t}_i$ for reproducing the bugs in $c_i$, \textit{i.e.},
\begin{equation}
    \mathbf{t}_i = \mathcal{L} \left(r_i, c_i, \mathcal{F}_i, \mathcal{T}_{\text{S}} \right),
    \label{eq: test_script_generate}
\end{equation}
where $\mathcal{T}_{\text{S}}$ denotes the test generation task template.
For a trustworthy evaluation, we further adopt an LLM-based judge agent to derive the semantic association between the revealed bugs and the generated tests, \textit{i.e.},
\begin{equation}
    \mathbf{M}_i =
    \mathcal{J}\left(r_i, c_i, \mathbf{t}_i, \mathcal{T}_{\text{J}}\right),
    \quad
    \mathbf{M}_i\in\{0,1\}^{|\mathbf{b}_i|\times |\mathbf{t}_i|},
    \label{eq: judge_matrix}
\end{equation}
where $\mathcal{J}$ denotes the judge agent, $\mathcal{T}_{\text{J}}$ denotes the judge task template, $\mathbf{b}_i$ denotes the set of bugs revealed in $c_i$, and $\mathbf{M}_i[j,k]=1$ indicates that the $j$-th bug in $\mathbf{b}_i$ is reproduced by the $k$-th test in $\mathbf{t}_i$.
Although the generated tests provide post-hoc evidence, they might still suffer from the following failure modes:
i) some tests might not exhibit fail-to-pass behavior and thus are invalid for reproducing bugs;
ii) even if all tests are valid, they might only cover a subset of the revealed bugs.
To capture these failure modes, we introduce the following metrics for more in-depth evaluation.

\textbf{Test Validity (TV):} we verify whether all generated tests exhibit fail-to-pass behavior, \textit{i.e.},
\begin{equation}
    \mathrm{TV} = \mathbb{I}\left[|\mathbf{t}_i|=|\mathbf{f}_i|\right],
\end{equation}
where $\mathbf{f}_i$ denotes the subset of generated tests that exhibit fail-to-pass behavior.

\textbf{Revealed:} we regard the revealed bugs as valid only if each of them is supported by test evidence, \textit{i.e.},
\begin{equation}
    \mathrm{Revealed} =
    \mathbb{I}
    \left[
    \forall j \in \{1,\ldots,|\mathbf{b}_i|\},\
    \exists k \in \{1,\ldots,|\mathbf{f}_i|\}:
    \mathbf{M}_i[j,k]=1
    \right].
\end{equation}
In the implementation, the Revealed metric is built upon $\mathrm{TV}=1$, thus ultimately indicating whether the agent reveals potential bugs with corresponding valid test evidence.

\section{Experiments}
In this section, we conduct extensive experiments on the proposed Active-SWE to evaluate the bug-fixing capability of state-of-the-art (SOTA) LLMs. 

\begin{figure*}[t]
\centering
\begin{minipage}[t]{0.42\textwidth}
    \vspace{0pt}
    \centering
    \includegraphics[width=0.78\linewidth]{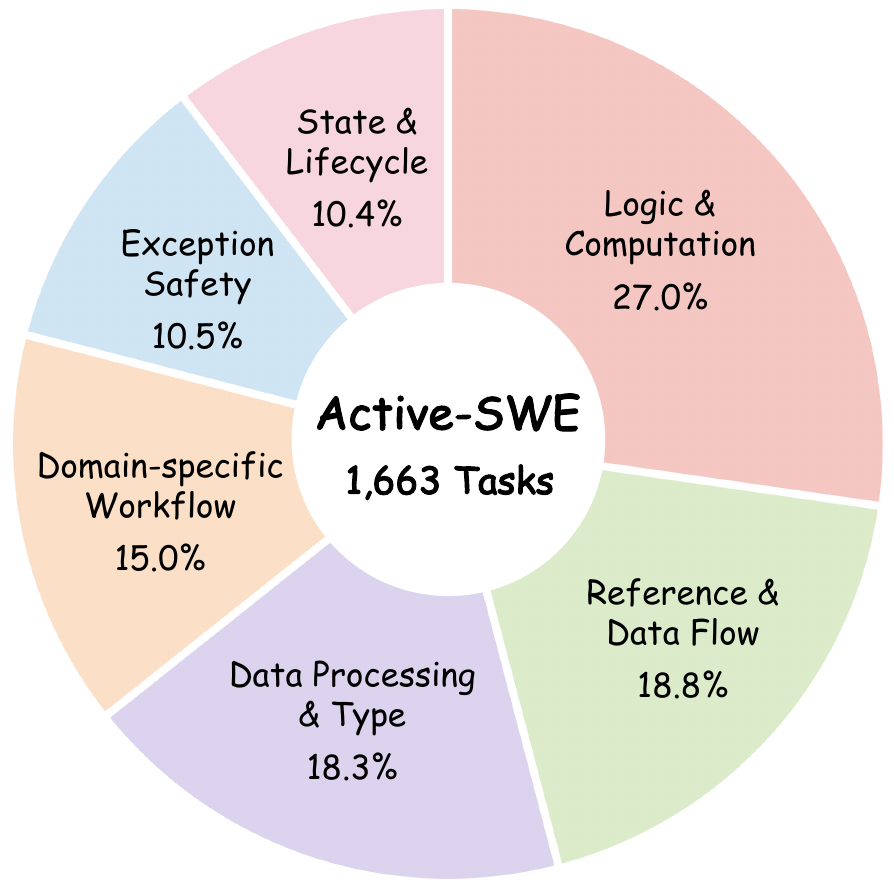}
    \captionof{figure}{
        Bug category distribution of the Active-SWE.
    }
    \label{fig: data_statistics}
\end{minipage}
\hspace{0.01\textwidth}
\begin{minipage}[t]{0.5\textwidth}
    \vspace{0pt}
    \centering
    \captionof{table}{
        Median values of different attributes in the Active-SWE under the simple and hard settings.
    }
    \label{tab: crbench_median_values}
    \begingroup
    \tablestyle{6pt}{1.2}
    \begin{tabular}{llcc}
        \toprule
        \textbf{Category} & \textbf{Attribute} & \textbf{Simple} & \textbf{Hard} \\
        \midrule
        \multirow{2}{*}{\textbf{Codebase}}
         & \# Lines & 193k & 353k \\
         & \# Files & 910 & 926 \\
        \midrule
        \multirow{3}{*}{\textbf{Gold Patch}}
         & \# Lines edited & 12 & 19 \\
         & \# Files edited & 1 & 2 \\
         & \# Hunks edited & 2 & 3 \\
        \midrule
        \multirow{2}{*}{\textbf{Tests}}
         & \# Fail to Pass & 1 & 2 \\
         & \# Pass to Pass & 16 & 104 \\
        \bottomrule
    \end{tabular}
    \endgroup
\end{minipage}
\end{figure*}

\subsection{Experiment Configurations}

\textbf{Data Statistics.}
For comprehensive evaluation, we construct 1,663 bug-fixing tasks with 1,411 simple instances and 252 hard instances, covering six major bug categories, \textit{i.e.}, Logic \& Computation, Reference \& Data Flow, Data Processing \& Type, Domain-specific Workflow, Exception Safety, and State \& Lifecycle and eight mainstream programming languages, \textit{i.e.}, Python, Go, Rust, PHP, Ruby, JavaScript$/$TypeScript (JS$/$TS), Java, and C$/$C++.
For high-quality and efficient evaluation, we curate Active-SWE with 400 tasks, including 300 simple and 100 hard instances.
To support broader-scale evaluation, we further construct Active-SWE-Extend with all constructed tasks.

\textbf{Baselines.}
To evaluate the bug-fixing capability of existing LLM-based coding agents, we select SOTA closed-source and open-source models with strong code reasoning capabilities, including Claude Opus 4.8~\citep{anthropicopus48} and Claude Sonnet 4.6~\citep{anthropic2026sonnet46}, GPT-5.5~\citep{gpt54} and GPT-5.4, Gemini-3.1-Pro~\citep{gemini31pro}, Qwen3.7-Max~\citep{qwen37}, Seed2.1~\citep{seed2}, DeepSeek-V4-Pro~\citep{deepseekv4}, GLM-5.2~\citep{glm5}, Kimi-K2.7-Code~\citep{kimik26} and Kimi-K2.6, MiniMax-M2.7~\citep{minimaxm27}, Ring-2.6-1T~\cite{ring2.6}, Hy3~\cite{hy3}, Qwen3.5 series~\citep{qwen35}.
For fairness, we adopt Claude Code as the unified scaffold for evaluation, as it is one of the most widely used agentic coding systems.

\begin{table*}[t]
\centering
\caption{Performance comparisons of different state-of-the-art LLMs.}
\label{tab: main_lite}
\tablestyle{11pt}{1.3}
\begin{tabular}{l|ccc|ccc}
\toprule
\multirow{2}{*}{\centering\textbf{Model}}& \multicolumn{3}{c|}{\textbf{Recorded}}
& \multicolumn{3}{c}{\textbf{Potential}} \\
& \textbf{LR}
& \textbf{LP}
& \textbf{Resolved}
& \textbf{Count}
& \textbf{TV}
& \textbf{Revealed} \\
\midrule
\rowcolor{gray!12}
\multicolumn{7}{c}{\textit{\textbf{Closed-Source Models}}} \\
\midrule
\modelicon{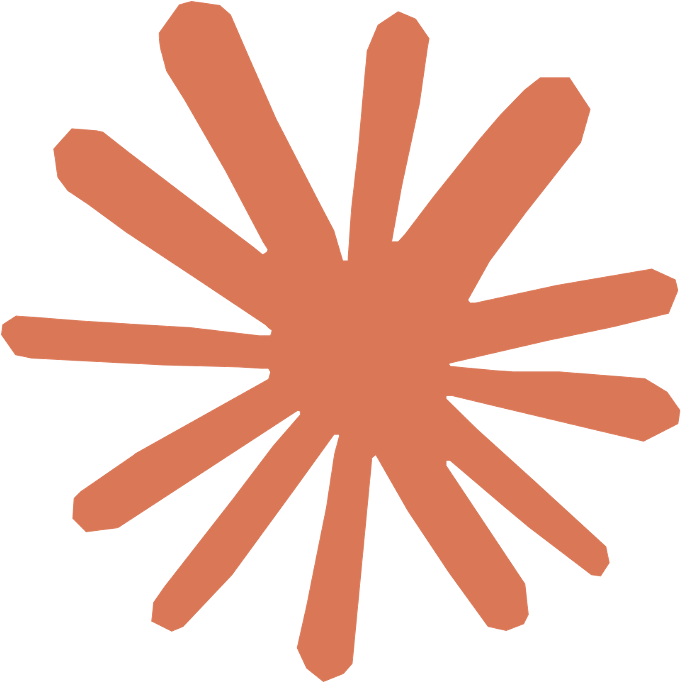}Claude Opus 4.8 & 29.1 & 35.0 & 20.0 & 1.2 & 81.0 & 75.0 \\
\modelicon{claude.png}Claude Sonnet 4.6 & 16.5 & 18.0 & 8.3& 1.6 & 78.3 & 53.8 \\
\modelicon{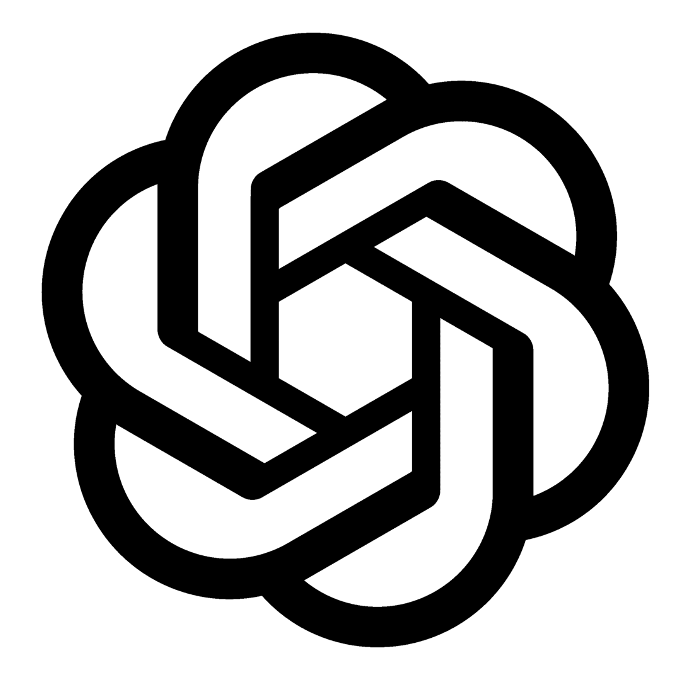}GPT-5.5 & 28.8 & 24.1 & 18.5 & 3.0 & 87.5 & 65.5 \\
\modelicon{gpt.png}GPT-5.4 & 17.2 & 17.6 & 8.8 & 1.3 & 52.3 & 32.8 \\
\modelicon{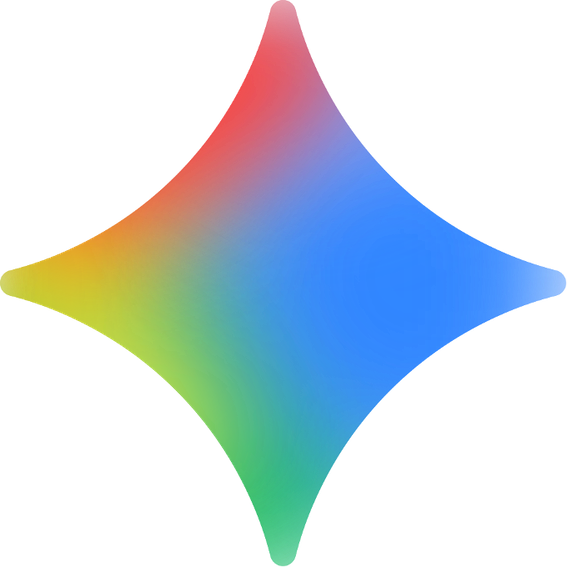}Gemini-3.1-Pro & 26.8 & 18.8 & 15.8 & 2.8 & 77.8 & 57.0 \\
\modelicon{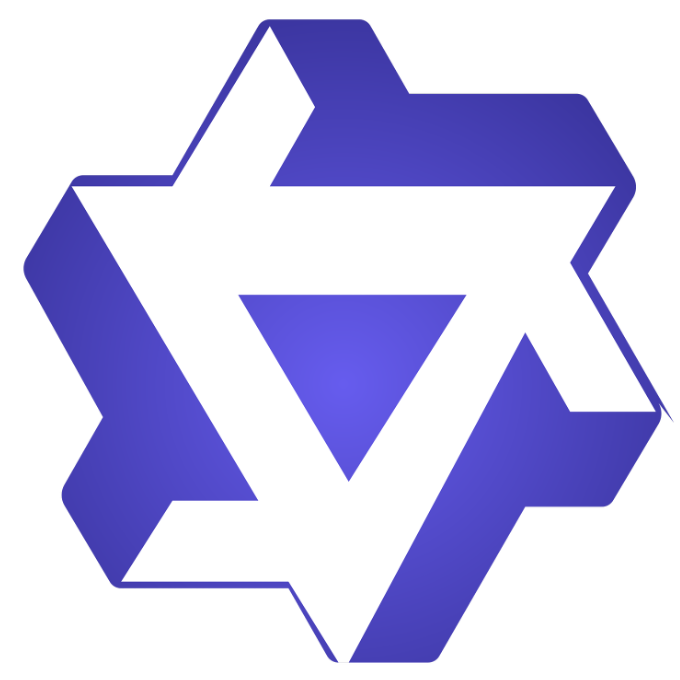}Qwen3.7-Max & 18.6 & 19.4 & 11.8 & 1.7 & 81.0 & 59.8 \\
\modelicon{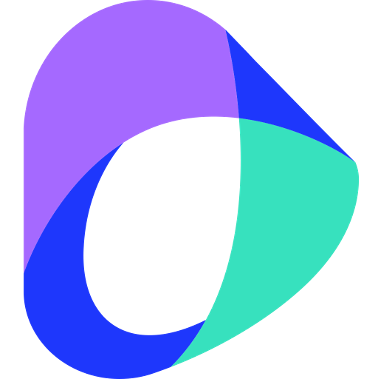}Seed2.1 & 21.1 & 20.3 & 11.0 & 1.8 & 64.5 & 52.8 \\
\midrule
\rowcolor{gray!12}
\multicolumn{7}{c}{\textit{\textbf{Open-Source Models}}} \\
\midrule
\modelicon{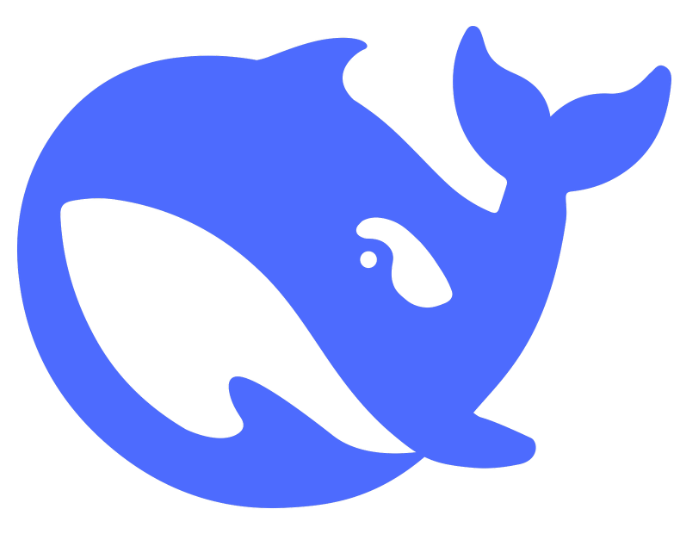}DeepSeek-V4-Pro & 17.5 & 18.9 & 9.5 & 2.0 & 64.3 & 53.3 \\
\modelicon{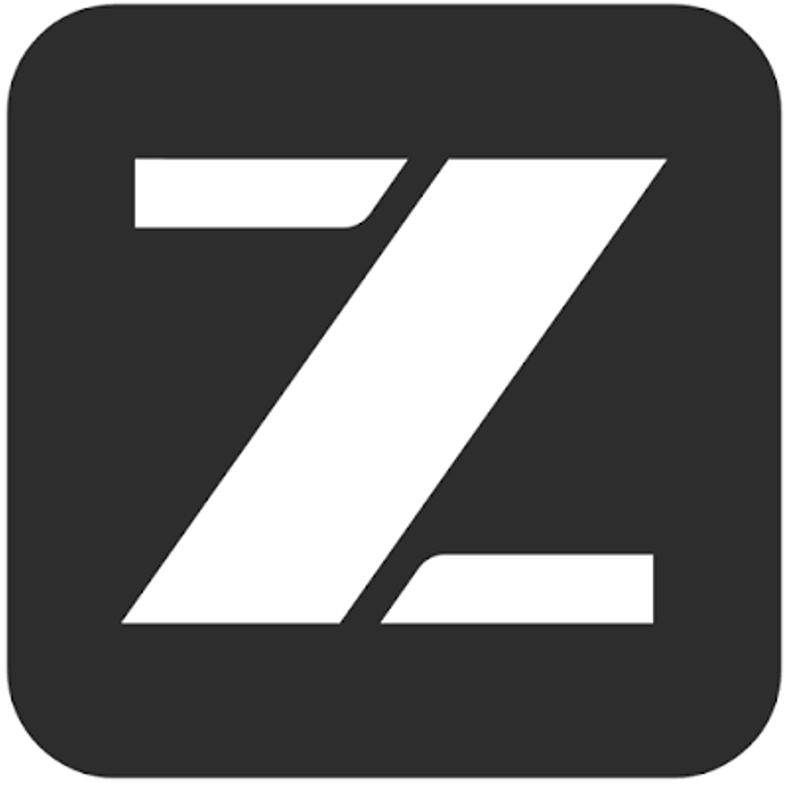}GLM-5.2 & 13.9 & 17.0 & 13.0 & 1.7 & 79.0 & 69.0 \\
\modelicon{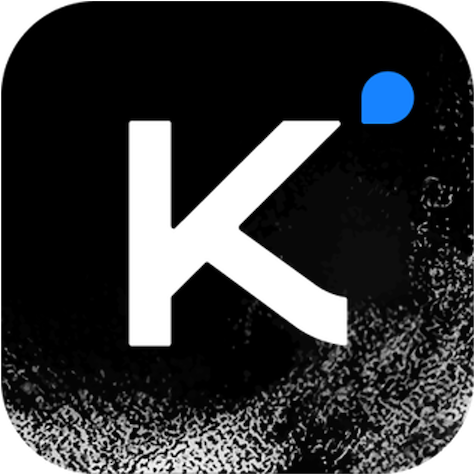}Kimi-K2.7-Code & 18.7 & 16.1 & 14.8 & 2.6 & 65.0 & 47.8 \\
\modelicon{kimi.png}Kimi-K2.6 & 17.2 & 14.7 & 14.0 & 2.1 & 67.3 & 51.8 \\
\modelicon{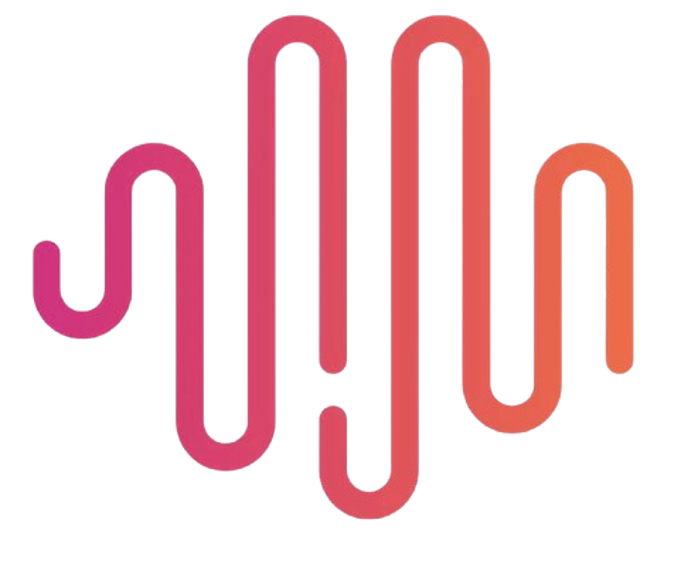}MiniMax-M2.7 & 9.9 & 13.4 & 7.8 & 1.6 & 45.0 & 38.5 \\
\modelicon{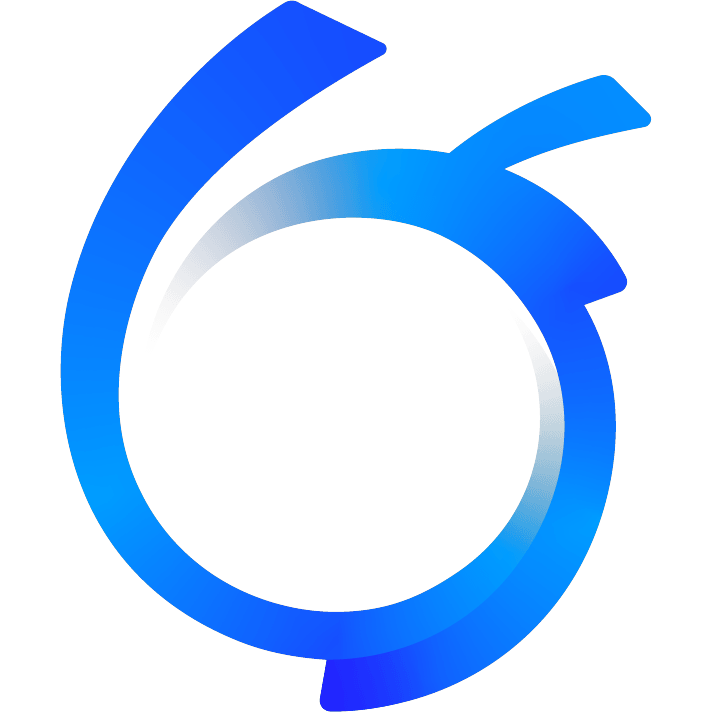}Ring-2.6-1T & 13.3 & 14.9 & 11.0 & 1.9 & 65.5 & 56.3 \\
\modelicon{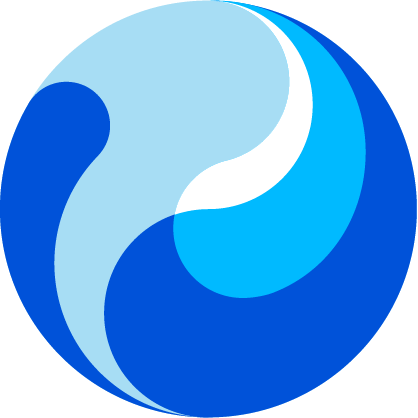}Hy3 & 12.3 & 15.8 & 9.8 & 1.1 & 77.3 & 58.5 \\
\modelicon{qwen.png}Qwen3.5-397B & 12.4 & 13.9 & 6.8 & 1.6 & 56.8 & 42.0 \\
\modelicon{qwen.png}Qwen3.5-122B & 11.2 & 13.8 & 5.8 & 1.4 & 46.8 & 40.0 \\
\modelicon{qwen.png}Qwen3.5-35B & 10.0 & 11.2 & 5.0 & 1.3 & 32.5 & 27.5 \\
\bottomrule
\end{tabular}
\end{table*}

\textbf{Implementation Details.}
We treat three closed-source models as LLM-based annotators for consensus-based taxonomy curation, including Gemini-3.1-Pro, GPT-5.4, and Claude Opus 4.8.
For potential bug evaluation, we adopt one of the most popular open-source models, \textit{i.e.}, Qwen3.5-397B, as the LLM-based judge agent to facilitate reproducible and accessible evaluation.
As for the hard setting, due to the high failure rate during temporal-aware integration, we set the window size to 2, \textit{i.e.}, each hard instance contains at least two recorded bugs.

\subsection{Main Results}
We compare SOTA LLMs on the Active-SWE benchmark in Table~\ref{tab: main_lite}, where ``Count'' refers to the number of generated tests.
From the results, one could have the following observations.
For recorded bug evaluation, existing LLMs achieve limited localization scores and resolved rates, with the best resolved rate reaching only 20.0\%, indicating that proactively discovering and repairing recorded bugs remains highly challenging.
Moreover, LLMs with stronger bug localization capability tend to achieve higher resolved rates, suggesting that localization quality is a precondition of repair success.
For potential bug evaluation, the ``Count'' and ``TV'' metrics are not always aligned, indicating that generating more bug-reproducing tests does not necessarily imply that these tests are valid.
Across various LLMs, Claude Opus 4.8 and GLM-5.2 achieve the strongest overall performance on both recorded bug fixing and potential bug discovery.
Besides, within the Qwen3.5 family, larger models with stronger coding capabilities tend to achieve better proactive bug-fixing performance.

\begin{figure*}[t]
\centering
\begin{minipage}[t]{0.55\textwidth}
    \vspace{0pt}
    \centering
    \includegraphics[width=0.9\linewidth]{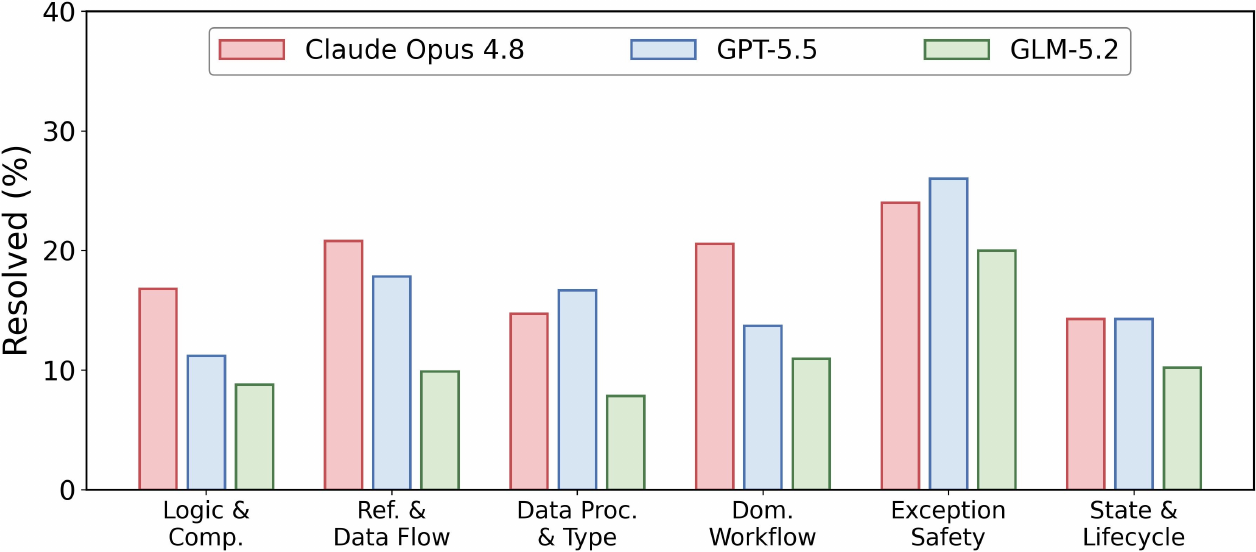}
    \captionof{figure}{
        Performance comparison across various bug categories.
    }
    \label{fig: various_bug}
\end{minipage}
\hspace{0.01\textwidth}
\begin{minipage}[t]{0.4\textwidth}
    \vspace{0pt}
    \centering
    \captionof{table}{
        Performance comparison between reactive and proactive bug fixing on a subset of Active-SWE.
    }
    \label{tab: reactive_proactive}
    \begingroup
    \tablestyle{2.3pt}{1.3}
    \begin{tabular}{l|cc}
        \toprule
        \multirow{2}{*}{\centering\textbf{Model}}
        & \multicolumn{2}{c}{\textbf{Resolved}} \\
        & \textbf{Reactive}
        & \textbf{Proactive} \\
        \midrule
        \modelicon{claude.png}Claude Opus 4.8 & 59.0 & 26.0 \\
        \modelicon{gpt.png}GPT-5.5 & 57.0 & 22.0 \\
        \modelicon{glm.png}GLM-5.2 & 58.0 & 17.0 \\
        \bottomrule
    \end{tabular}
    \endgroup
\end{minipage}
\end{figure*}

\begin{figure*}[t]
    \centering

    \includegraphics[width=0.95\linewidth]{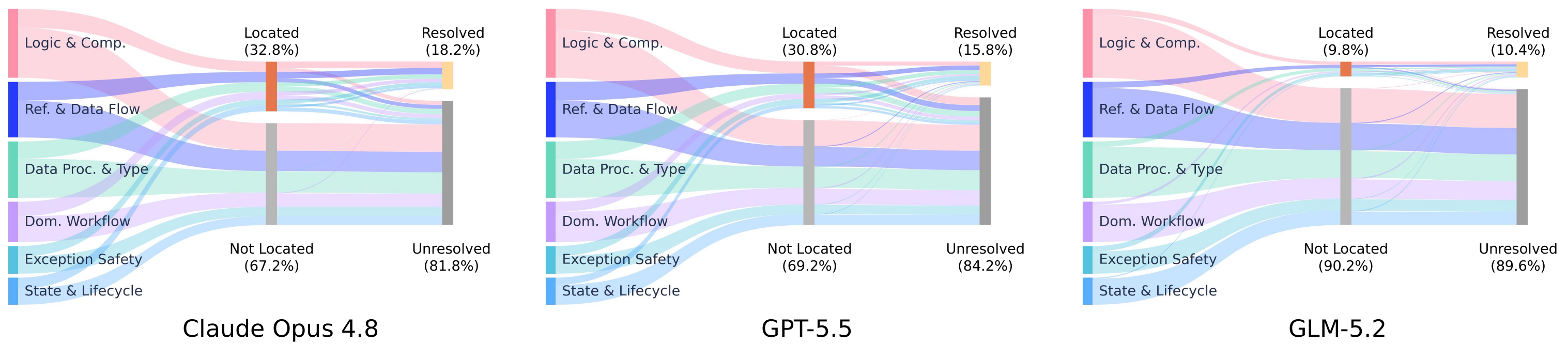}
    \caption{
    Relationship between bug localization and resolution. The recorded bugs are considered as \textit{Located} if $\mathrm{LR}\geq 0.5$.
    }
    \label{fig: find_resolve}

    \vspace{1.2em}

    \includegraphics[width=0.95\linewidth]{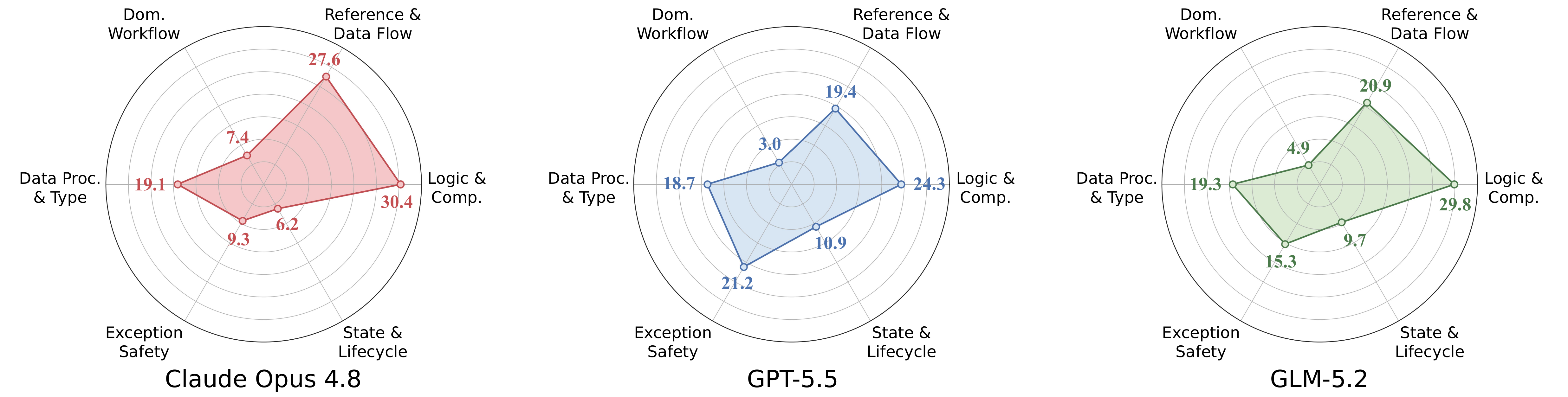}
    \caption{
    Distribution of revealed potential bugs.
    }
    \label{fig: revealed}
\end{figure*}

\subsection{Analytic Experiments}

\textbf{Proactive bug-fixing capability varies across bug categories.}
As shown in Fig.~\ref{fig: various_bug}, LLMs achieve better resolved rates on Exception Safety and struggle with bugs of State \& Lifecycle, suggesting that LLMs are better at discovering and fixing certain categories of bugs.

\textbf{Proactive bug fixing is more challenging than reactive bug fixing.}
As demonstrated in Table~\ref{tab: reactive_proactive}, LLMs suffer performance degradation in the proactive setting compared with the reactive setting, suggesting that even SOTA LLMs still struggle to resolve bugs without external guidance.

\textbf{Right bug localization is halfway to successful bug fixing.}
From the results in Fig.~\ref{fig: find_resolve}, correctly located bugs are more likely to be resolved, suggesting that accurate localization is positively associated with successful repair.
It is worth noting that some successful repairs do not match the reference oracle, suggesting that LLMs may discover alternative repair paths beyond the human-recorded one.

\textbf{LLMs tend to reveal certain potential bugs.}
We visualize the category distribution of revealed potential bugs in Fig.~\ref{fig: revealed}.
The revealed bugs are concentrated in Logic \& Computation, Reference \& Data Flow, with fewer cases in Domain-specific Workflow and State \& Lifecycle.
Such a phenomenon suggests that LLMs are more sensitive to discovering local correctness and flow-related issues than to domain- or state-dependent bugs.

\begin{figure*}[t]
    \centering
    \includegraphics[width=0.95\linewidth]{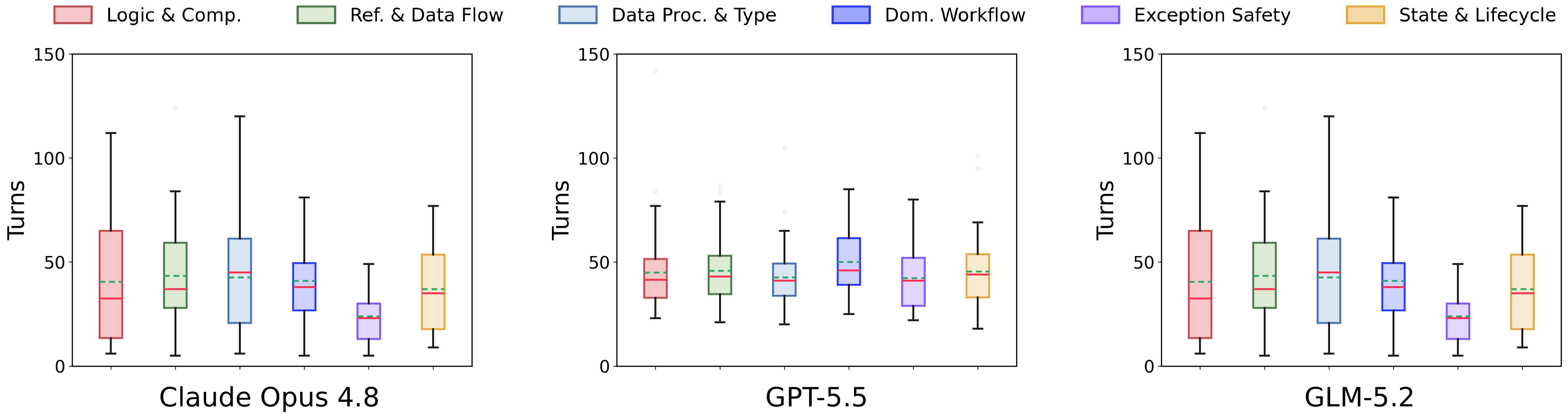}
    \caption{
    Reasoning turns across various recorded bug categories.
    }
    \label{fig: turns}
\end{figure*}

\textbf{LLMs exhibit similar reasoning depth across bug categories.}
As demonstrated in Fig.~\ref{fig: turns}, reasoning turns are broadly comparable across bug categories within each model, indicating that LLMs exhibit similar reasoning depth and interaction costs regardless of the target bug categories.

\begin{figure*}[t]
    \centering
    \includegraphics[width=0.95\linewidth]{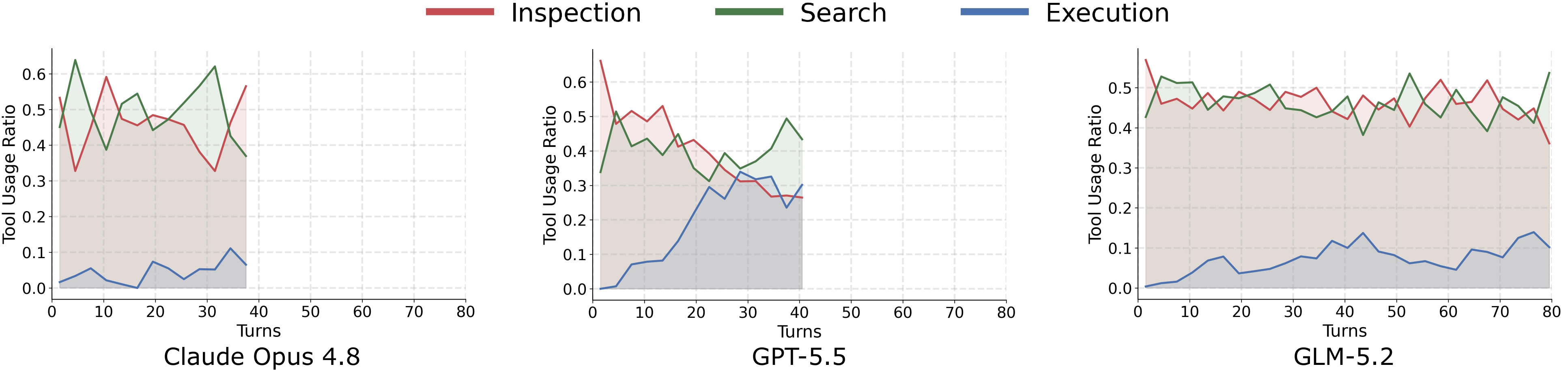}
    \caption{
    Tool-usage patterns and turn dynamics in proactive bug fixing.
    }
    \label{fig: issue_free}
\end{figure*}

\textbf{Different LLMs share common reasoning patterns but differ in reasoning depth.}
From the results in Fig.~\ref{fig: issue_free}, LLM-based coding agents invest more early-stage effort in code inspection in the proactive scenario and then shift from inspection to search and execution, highlighting a progressive process from repository exploration to bug localization and repair.
Besides, different LLMs exhibit distinct reasoning depths, \textit{e.g.}, GLM-5.2 allocates more turns for in-depth analysis, while Claude Opus 4.8 resolves bugs with shorter and more efficient trajectories.

\section{Conclusion}
In this paper, we propose Active-SWE, a novel benchmark for comprehensively evaluating LLM-based coding agents in proactive bug-fixing capability.
With the proposed data curation pipeline, Active-SWE curates 1,663 high-quality bug-fixing tasks spanning six major bug categories and eight programming languages.
To facilitate more in-depth validation, we design a novel task formulation strategy and a dual-track evaluation framework, which expand the evaluation scope from both difficulty and coverage perspectives.
Extensive experiments reveal the limitations of existing SOTA LLMs in proactive bug fixing and provide insights for subsequent research.
In the future, we plan to explore broader downstream scenarios of proactive bug fixing, such as vulnerability discovery and security-oriented code auditing.



\bibliography{iclr2025_conference}

@inproceedings{swe-bench,
  title={Swe-bench: Can language models resolve real-world github issues?},
  author={Jimenez, Carlos E and Yang, John and Wettig, Alexander and Yao, Shunyu and Pei, Kexin and Press, Ofir and Narasimhan, Karthik},
  booktitle={ICLR},
  year={2024}
}

@article{swe-bench-pro,
  title={Swe-bench pro: Can ai agents solve long-horizon software engineering tasks?},
  author={Deng, Xiang and Da, Jeff and Pan, Edwin and He, Yannis Yiming and Ide, Charles and Garg, Kanak and Lauffer, Niklas and Park, Andrew and Pasari, Nitin and Rane, Chetan and others},
  journal={arXiv preprint arXiv:2509.16941},
  year={2025}
}

@article{llm1,
  title={A survey on evaluation of large language models},
  author={Chang, Yupeng and Wang, Xu and Wang, Jindong and Wu, Yuan and Yang, Linyi and Zhu, Kaijie and Chen, Hao and Yi, Xiaoyuan and Wang, Cunxiang and Wang, Yidong and others},
  journal={ACM transactions on intelligent systems and technology},
  year={2024}
}

@article{llm2,
  title={A survey of large language models},
  author={Zhao, Wayne Xin and Zhou, Kun and Li, Junyi and Tang, Tianyi and Dong, Zican and Hou, Yupeng and Zhang, Beichen and Min, Yingqian and Zhang, Junjie and Liu, Peiyu and others},
  journal={Frontiers of Computer Science},
  year={2026}
}

@inproceedings{swe1,
  title={Large language models for software engineering: Survey and open problems},
  author={Fan, Angela and Gokkaya, Beliz and Harman, Mark and Lyubarskiy, Mitya and Sengupta, Shubho and Yoo, Shin and Zhang, Jie M},
  booktitle={ICSE-FoSE},
  year={2023}
}

@article{swe2,
  title={From llms to llm-based agents for software engineering: A survey of current, challenges and future},
  author={Jin, Haolin and Huang, Linghan and Cai, Haipeng and Yan, Jun and Li, Bo and Chen, Huaming},
  journal={arXiv preprint arXiv:2408.02479},
  year={2024}
}

@misc{onehand_case1,
  title       = {External Technical Root Cause Analysis — Channel File 291},
  author      = {{CrowdStrike}},
  year        = {2024},
  howpublished = {\url{https://www.crowdstrike.com/wp-content/uploads/2024/08/Channel-File-291-Incident-Root-Cause-Analysis-08.06.2024.pdf}}
}

@misc{onehand_case2,
  title       = {Helping our customers through the CrowdStrike outage},
  author      = {{Microsoft}},
  year        = {2024},
  howpublished = {\url{https://blogs.microsoft.com/blog/2024/07/20/helping-our-customers-through-the-crowdstrike-outage/}}
}

@misc{onehand_case3,
  title       = {CrowdStrike to Cost Fortune 500 \$5.4b; Insured Loss Range of 
  \$0.54b - \$1.08b},
  author      = {{Parametrix}},
  year        = {2024},
  howpublished = {\url{https://www.parametrixinsurance.com/in-the-news/crowdstrike-to-cost-fortune-500-5-4-billion-insured-loss-range-of-540-million-to-1-08-billion}}
}

@misc{otherhand_case1,
  title       = {Why SWE-bench Verified no longer measures frontier coding capabilities},
  author      = {{OpenAI}},
  year        = {2026},
  howpublished = {\url{https://openai.com/index/why-we-no-longer-evaluate-swe-bench-verified/}}
}

@misc{otherhand_case2,
  title       = {Separating signal from noise in coding evaluations},
  author      = {{OpenAI}},
  year        = {2026},
  howpublished = {\url{https://openai.com/index/separating-signal-from-noise-coding-evaluations/}}
}

@inproceedings{swe-bench-live,
  title={Swe-bench goes live!},
  author={Zhang, Linghao and He, Shilin and Zhang, Chaoyun and Kang, Yu and Li, Bowen and Xie, Chengxing and Wang, Junhao and Wang, Maoquan and Huang, Yufan and Fu, Shengyu and others},
  booktitle={NeurIPS},
  year={2026}
}

@article{swe-bench-multimodal,
  title={Swe-bench multimodal: Do ai systems generalize to visual software domains?},
  author={Yang, John and Jimenez, Carlos E and Zhang, Alex L and Lieret, Kilian and Yang, Joyce and Wu, Xindi and Press, Ori and Muennighoff, Niklas and Synnaeve, Gabriel and Narasimhan, Karthik R and others},
  journal={arXiv preprint arXiv:2410.03859},
  year={2024}
}

@inproceedings{multi-swe-bench,
  title={Multi-swe-bench: A multilingual benchmark for issue resolving},
  author={Zan, Daoguang and Huang, Zhirong and Liu, Wei and Chen, Hanwu and Xin, Shulin and Zhang, Linhao and Liu, Qi and Aoyan, Li and Chen, Lu and Zhong, Xiaojian and others},
  booktitle={NeurIPS},
  year={2026}
}

@article{featurebench,
  title={Featurebench: Benchmarking agentic coding for complex feature development},
  author={Zhou, Qixing and Zhang, Jiacheng and Wang, Haiyang and Hao, Rui and Wang, Jiahe and Han, Minghao and Yang, Yuxue and Wu, Shuzhe and Pan, Feiyang and Fan, Lue and others},
  journal={arXiv preprint arXiv:2602.10975},
  year={2026}
}

@article{swe-atlas,
  title={SWE Atlas: Benchmarking Coding Agents Beyond Issue Resolution},
  author={Raghavendra, Mohit and Dan, Soham and Calvo, Miguel Romero and He, Yannis Yiming and Mols, Johannes Baptist and Anand, Gautam and McCollum, Cole and Arakelyan, Edgar and Bharadwaj, Vijay and Park, Andrew and others},
  journal={arXiv preprint arXiv:2605.08366},
  year={2026}
}

@article{testexplora,
  title={TestExplora: Benchmarking LLMs for Proactive Bug Discovery via Repository-Level Test Generation},
  author={Liu, Steven and Luo, Jane and Zhang, Xin and Liu, Aofan and Liu, Hao and Wu, Jie and Huang, Ziyang and Huang, Yangyu and Kang, Yu and Li, Scarlett},
  journal={arXiv preprint arXiv:2602.10471},
  year={2026}
}

@inproceedings{swt-bench,
  title={Swt-bench: Testing and validating real-world bug-fixes with code agents},
  author={M{\"u}ndler, Niels and M{\"u}ller, Mark N and He, Jingxuan and Vechev, Martin},
  booktitle={NeurIPS},
  year={2024}
}

@article{codeagent1,
  title={Agentcoder: Multi-agent-based code generation with iterative testing and optimisation},
  author={Huang, Dong and Zhang, Jie M and Luck, Michael and Bu, Qingwen and Qing, Yuhao and Cui, Heming},
  journal={arXiv preprint arXiv:2312.13010},
  year={2023}
}

@article{swe-agent,
  title={Swe-agent: Agent-computer interfaces enable automated software engineering},
  author={Yang, John and Jimenez, Carlos and Wettig, Alexander and Lieret, Kilian and Yao, Shunyu and Narasimhan, Karthik and Press, Ofir},
  journal={Advances in Neural Information Processing Systems},
  volume={37},
  pages={50528--50652},
  year={2024}
}

@inproceedings{codeagent2,
  title={Codeagent: Enhancing code generation with tool-integrated agent systems for real-world repo-level coding challenges},
  author={Zhang, Kechi and Li, Jia and Li, Ge and Shi, Xianjie and Jin, Zhi},
  booktitle={ACL},
  year={2024}
}

@inproceedings{openhands,
  title={Openhands: An open platform for ai software developers as generalist agents},
  author={Wang, Xingyao and Li, Boxuan and Song, Yufan and Xu, Frank F and Tang, Xiangru and Zhuge, Mingchen and Pan, Jiayi and Song, Yueqi and Li, Bowen and Singh, Jaskirat and others},
  booktitle={ICLR},
  year={2025}
}

@inproceedings{codeagent3,
  title={Mapcoder: Multi-agent code generation for competitive problem solving},
  author={Islam, Md Ashraful and Ali, Mohammed Eunus and Parvez, Md Rizwan},
  booktitle={ACL},
  year={2024}
}

@inproceedings{codeagent_swe1,
  title={Swe-smith: Scaling data for software engineering agents},
  author={Yang, John and Lieret, Kilian and Jimenez, Carlos and Wettig, Alexander and Khandpur, Kabir and Zhang, Yanzhe and Hui, Binyuan and Press, Ofir and Schmidt, Ludwig and Yang, Diyi},
  booktitle={NeurIPS},
  year={2026}
}

@article{codeagent_swe2,
  title={Live-SWE-agent: Can Software Engineering Agents Self-Evolve on the Fly?},
  author={Xia, Chunqiu Steven and Wang, Zhe and Yang, Yan and Wei, Yuxiang and Zhang, Lingming},
  journal={arXiv preprint arXiv:2511.13646},
  year={2025}
}

@article{react,
  title={React: Synergizing reasoning and acting in language models},
  author={Yao, Shunyu and Zhao, Jeffrey and Yu, Dian and Du, Nan and Shafran, Izhak and Narasimhan, Karthik and Cao, Yuan},
  journal={arXiv preprint arXiv:2210.03629},
  year={2022}
}

@misc{anthropic2026sonnet46,
  title        = {System Card: Claude Sonnet 4.6},
  author       = {{Anthropic}},
  year         = {2026},
  howpublished = {\url{https://www-cdn.anthropic.com/78073f739564e986ff3e28522761a7a0b4484f84.pdf}},
}

@article{gpt54,
  title={Openai gpt-5 system card},
  author={Singh, Aaditya and Fry, Adam and Perelman, Adam and Tart, Adam and Ganesh, Adi and El-Kishky, Ahmed and McLaughlin, Aidan and Low, Aiden and Ostrow, AJ and Ananthram, Akhila and others},
  journal={arXiv preprint arXiv:2601.03267},
  year={2025}
}

@misc{deepseekv4,
  title        = {DeepSeek-V4: Towards Highly Efficient Million-Token Context Intelligence},
  author       = {{DeepSeek-AI}},
  year         = {2026},
  howpublished = {\url{https://huggingface.co/deepseek-ai/DeepSeek-V4-Pro/blob/main/DeepSeek_V4.pdf}}
}

@article{kimik26,
  title={Kimi K2. 5: Visual Agentic Intelligence},
  author={Team, Kimi and Bai, Tongtong and Bai, Yifan and Bao, Yiping and Cai, SH and Cao, Yuan and Charles, Y and Che, HS and Chen, Cheng and Chen, Guanduo and others},
  journal={arXiv preprint arXiv:2602.02276},
  year={2026}
}

@article{minimaxm27,
  title={The minimax-m2 series: Mini activations unleashing max real-world intelligence},
  author={Chen, Aili and Li, Aonian and Zhou, Baichuan and Gong, Bangwei and Jiang, Binyang and Dan, Boji and Yu, Changqing and Wang, Chao and Ma, Cheng and Zhong, Cheng and others},
  journal={arXiv preprint arXiv:2605.26494},
  year={2026}
}

@misc{qwen35,
    title = {Qwen3.5: Accelerating Productivity with Native Multimodal Agents},
    howpublished = {\url{https://qwen.ai/blog?id=qwen3.5}},
    author = {Qwen Team},
    year = {2026}
}

@misc{gemini31pro,
  title        = {Gemini 3.1 Pro Model Card},
  author       = {{Google DeepMind}},
  year         = {2026},
  howpublished = {\url{https://deepmind.google/models/model-cards/gemini-3-1-pro/}}
}

@misc{anthropicopus48,
  title        = {System Card: Claude Opus 4.8},
  author       = {{Anthropic}},
  year         = {2026},
  howpublished = {\url{https://www-cdn.anthropic.com/0b4915911bb0d19eca5b5ee635c80fef830a37ea.pdf}}
}

@article{glm5,
  title={Glm-5: from vibe coding to agentic engineering},
  author={Zeng, Aohan and Lv, Xin and Hou, Zhenyu and Du, Zhengxiao and Zheng, Qinkai and Chen, Bin and Yin, Da and Ge, Chendi and Huang, Chenghua and Xie, Chengxing and others},
  journal={arXiv preprint arXiv:2602.15763},
  year={2026}
}

@misc{qwen37,
    title = {{Qwen3.7}: The Agent Frontier},
    url = {https://qwen.ai/blog?id=qwen3.7},
    author = {{Qwen Team}},
    month = {May},
    year = {2026}
}

@article{seed2,
  title={Seed2. 0 model card: Towards intelligence frontier for real-world complexity},
  author={Seed, Bytedance},
  journal={arXiv preprint arXiv:2607.00248},
  year={2026}
}

@article{ring2.6,
  title={Ling and Ring 2.6 Technical Report: Efficient and Instant Agentic Intelligence at Trillion-Parameter Scale},
  author={Li, Ang and Liu, Ben and Han, Bin and Hu, Bin and Jing, Bin and Hu, Binbin and Li, Bing and Chen, Cai and Tang, Caizhi and Tian, Changxin and others},
  journal={arXiv preprint arXiv:2606.15079},
  year={2026}
}

@misc{hy3,
  author       = {{Tencent Hy Team}},
  title        = {{Hy3}},
  year         = {2026},
  howpublished = {\url{https://huggingface.co/tencent/Hy3}}
}
\bibliographystyle{iclr2025_conference}

\newpage
\appendix
\begin{leftline}
	{
		\LARGE{\textsc{Appendix}}
	}
\end{leftline}

    \etocdepthtag.toc{mtappendix}
    \etocsettagdepth{mtchapter}{none}
    \etocsettagdepth{mtappendix}{subsection}
    \etocsettocstyle{}{}
    
    {
        \hypersetup{linkcolor=black}
        \footnotesize\tableofcontents
    }

\newpage

\section{More Data Statistics}
In this section, we present additional data statistics of our benchmark in Tables~\ref{tab: feature_distribution_sweactiveextend} and~\ref{tab: feature_distribution_sweactive}, which report the feature distributions of Active-SWE-Extend and the curated Active-SWE.

Active-SWE-Extend exhibits variation in patch complexity across languages.
Specifically, Rust and Go involve more fragmented patches, with 12.2 and 9.5 hunks on average, respectively, while Go and Java touch the largest number of files.
In contrast, the code patches of PHP and Ruby instances are generally more localized, involving fewer files and hunks per patch.
Besides, the test statistics also vary notably, \textit{e.g.}, Python and Ruby have the largest average number of pass-to-pass tests, while JS$/$TS and Python have the highest average number of fail-to-pass tests.

Active-SWE preserves the multilingual and heterogeneous nature of Active-SWE-Extend and provides a compact subset for main experiments, which includes 400 instances across the same eight languages.
Compared with Active-SWE-Extend, Active-SWE retains diverse patch and test characteristics, including localized edits in Ruby and PHP, more fragmented patches in Go and Rust, and non-trivial regression coverage in Python and Ruby, and so on.
These statistics show that the subset remains representative enough for evaluating realistic bug-fixing capabilities while reducing evaluation cost.

\begin{table*}[htbp]
\centering
\caption{Feature distribution of Active-SWE-Extend instances across programming languages.}
\label{tab: feature_distribution_sweactiveextend}
\tablestyle{8pt}{0.9}
\begin{tabular}{lccccc}
\toprule
& \multicolumn{1}{c}{Instance}
& \multicolumn{2}{c}{Code Patch}
& \multicolumn{2}{c}{Unit Test} \\
\cmidrule(lr){2-2}
\cmidrule(lr){3-4}
\cmidrule(lr){5-6}
Language
& \# Num
& \# Files
& \# Hunks
& \# F2P
& \# P2P \\
\midrule
Python & 505 & 1.8 & 3.3 & 2.2 & 114.2 \\
Go & 327 & 2.2 & 9.5 & 1.3 & 26.8 \\
Rust & 153 & 1.7 & 12.2 & 1.5 & 28.9 \\
PHP & 250 & 1.7 & 3.0 & 1.3 & 40.5 \\
Ruby & 98 & 1.7 & 3.0 & 1.8 & 67.9 \\
JS/TS & 181 & 2.1 & 4.5 & 3.2 & 28.7 \\
Java & 76 & 2.2 & 4.2 & 1.1 & 0.2 \\
C/C++ & 73 & 1.7 & 7.7 & 1.6 & 42.8 \\
\bottomrule
\end{tabular}
\end{table*}

\begin{table*}[htbp]
\centering
\caption{Feature distribution of Active-SWE instances across programming languages.}
\label{tab: feature_distribution_sweactive}
\tablestyle{8pt}{0.9}
\begin{tabular}{lccccc}
\toprule
& \multicolumn{1}{c}{Instance}
& \multicolumn{2}{c}{Code Patch}
& \multicolumn{2}{c}{Unit Test} \\
\cmidrule(lr){2-2}
\cmidrule(lr){3-4}
\cmidrule(lr){5-6}
Language
& \#Num
& \# Files
& \# Hunks
& \# F2P
& \# P2P \\
\midrule
Python & 103 & 2.1 & 3.2 & 3.5 & 118.9 \\
Go & 59 & 2.1 & 6.0 & 1.4 & 8.8 \\
Rust & 45 & 2.3 & 5.5 & 1.6 & 9.8 \\
PHP & 42 & 1.7 & 3.1 & 1.5 & 13.2 \\
Ruby & 29 & 1.4 & 2.1 & 1.8 & 62.9 \\
JS/TS & 65 & 2.2 & 4.4 & 3.3 & 26.6 \\
Java & 27 & 2.1 & 4.3 & 1.0 & 0.0 \\
C/C++ & 30 & 1.9 & 4.4 & 1.9 & 32.6 \\
\bottomrule
\end{tabular}
\end{table*}

\newpage

\section{Discussions on Real-world SWE Application}
\label{appendix:discussions_swe_application}
As discussed in the manuscript, Active-SWE focus on evaluating proactive bug fixing capabilities within files pending review.
Such a setting corresponds to a widely-used development workflow, where developers review recently modified files before opening a pull request.
It is worth noting that the design of review scope could also serve as a subtask of proactive bug fixing for the whole repository.
In other words, our setting naturally supports this broader scenario by decomposing the repository-level task into successive file-level tasks that collectively cover the codebase.

To further verify this claim, we conduct additional experiments on two stable and mature repositories, \textit{i.e.}, the pflag and Click repositories, each containing ten recorded bugs as verifiable evaluation targets.
Specifically, we compare two settings using SOTA coding agent, \textit{i.e.}, GLM-5.2, to verify the rationale for our design:
i) \textit{global review}, which instructs the coding agent to proactively fix bugs across the entire repository in a single run;
ii) \textit{iterative local review}, which requires the coding agent to review a few source files, \textit{e.g.}, two files, at a time until the entire repository is covered.

From the results in Table~\ref{tab:repository_review}, simply performing global review across the entire repository barely identifies historically recorded bugs, while our design, \textit{i.e.}, iterative local review, substantially improves bug localization and resolved rates.
Such results suggest that existing coding agents struggle to find and resolve recorded bugs when directly asked to review the entire repository.
In contrast, the formulated task evaluate proactive bug fixing capability within specified review scope and could be extended to the scope of the whole repository by decomposing global review into successive local review tasks.

\begin{table*}[htbp]
    \centering
    \caption{Comparison of global review and iterative local review on recorded bug fixing.}
    \label{tab:repository_review}
    \tablestyle{8pt}{1.15}
    \begin{tabular}{ll|ccc}
        \toprule
        \textbf{Repository} & \textbf{Setting}
        & \textbf{LR} & \textbf{LP} & \textbf{Resolved} \\
        \midrule
        \multirow{2}{*}{pflag}
        & Global review & 0.0 & 0.0 & 0.0 \\
        & Iterative local review & 10.6 & 23.8 & 20.0 \\
        \midrule
        \multirow{2}{*}{Click}
        & Global review & 0.0 & 0.0 & 0.0 \\
        & Iterative local review & 4.7 & 33.3 & 10.0 \\
        \bottomrule
    \end{tabular}
\end{table*}

\newpage

\section{More Experimental Results}

\subsection{Experimental Results under Simple and Hard Settings}
In the main manuscript, we have conducted comprehensive experiments on curated Active-SWE, which contains 400 proactive bug-fixing tasks.
Here, we provide more experimental results on the 1,663 tasks in Active-SWE-Extend.
Specifically, for reproducible and accessible evaluation, we evaluate several open-source models on Active-SWE-Extend in Table~\ref{tab: simple_setting}-\ref{tab: hard_setting}.

From the results, one could have the following observations and conclusions:
i) for recorded bug fixing, LLMs still achieve limited localization scores and resolved rates under the simple and hard settings, suggesting that LLMs struggle with both needle-in-a-haystack bug localization and subsequent repair generation;
ii) for potential bug discovery, although the review scope contains a few recorded bugs, LLMs are able to reveal additional potential bugs;
iii) compared with the simple setting that involves one recorded bug, LLMs rarely resolve multiple recorded bugs simultaneously under the hard setting, further highlighting the difficulty of comprehensive proactive bug fixing;
iv) under the hard setting, even when LLMs fail to resolve all recorded bugs, they may fix a subset of them, suggesting that LLMs tend to perform incomplete bug fixing.

\begin{table*}[ht]
\centering
\caption{Performance Comparisons of different SOTA models under the simple setting.}
\label{tab: simple_setting}
\tablestyle{11pt}{1.3}
\begin{tabular}{l|ccc|ccc}
\toprule
\multirow{2}{*}{\centering\textbf{Model}}& \multicolumn{3}{c|}{\textbf{Recorded}}
& \multicolumn{3}{c}{\textbf{Potential}} \\
& \textbf{LR}
& \textbf{LP}
& \textbf{Resolved}
& \textbf{Count}
& \textbf{TV}
& \textbf{Revealed} \\
\midrule
\modelicon{deepseek.png}DeepSeek-V4-Pro & 15.2 & 16.0 & 9.4 & 1.7 & 77.0 & 61.3 \\
\modelicon{kimi.png}Kimi-K2.6 & 17.8 & 14.6 & 13.8 & 2.1 & 73.8 & 56.5 \\
\modelicon{minimax.png}MiniMax-M2.7 & 10.5 & 13.6 & 5.1 & 1.5 & 42.9 & 36.2 \\
\modelicon{qwen.png}Qwen3.5-397B & 11.5 & 14.6 & 6.2 & 1.5 & 55.7 & 44.6 \\
\modelicon{qwen.png}Qwen3.5-122B & 11.4 & 14.0 & 6.0 & 1.5 & 48.1 & 38.5 \\
\modelicon{qwen.png}Qwen3.5-35B & 8.6 & 11.0 & 5.2 & 1.4 & 28.8 & 22.7 \\
\bottomrule
\end{tabular}
\end{table*}

\begin{table*}[ht]
\centering
\caption{Performance Comparisons of different SOTA models under the Hard setting.}
\label{tab: hard_setting}
\tablestyle{8pt}{1.3}
\begin{tabular}{l|cccc|ccc}
\toprule
\multirow{2}{*}{\centering\textbf{Model}}
& \multicolumn{4}{c|}{\textbf{Recorded}}
& \multicolumn{3}{c}{\textbf{Potential}} \\
& \textbf{LR}
& \textbf{LP}
& \textbf{Res.}
& \textbf{Sub Res.}
& \textbf{Count}
& \textbf{TV}
& \textbf{Revealed} \\
\midrule
\modelicon{deepseek.png}DeepSeek-V4-Pro   & 7.9 & 9.5 & 0.4 & 5.2 & 2.2 & 82.1 & 62.3 \\
\modelicon{kimi.png}Kimi-K2.6     & 10.4 & 8.1 & 1.2 & 6.5 & 2.9 & 69.0 & 56.7 \\
\modelicon{minimax.png}MiniMax-M2.7  & 3.7 & 6.8 & 0.4 & 2.0 & 1.9 & 55.2 & 45.6 \\
\modelicon{qwen.png}Qwen3.5-397B  & 5.4 & 7.6 & 0.8 & 3.2 & 1.5 & 60.3 & 46.8 \\
\modelicon{qwen.png}Qwen3.5-122B  & 3.5 & 5.7 & 0.4 & 2.2 & 2.0 & 48.0 & 36.5 \\
\modelicon{qwen.png}Qwen3.5-35B   & 5.1 & 7.2 & 0.8 & 2.6 & 2.2 & 29.8 & 17.1 \\
\bottomrule
\end{tabular}
\end{table*}

\newpage

\subsection{Experimental Results across Programming Languages}
\begin{figure*}[htbp]
    \centering
    \includegraphics[width=0.95\linewidth]{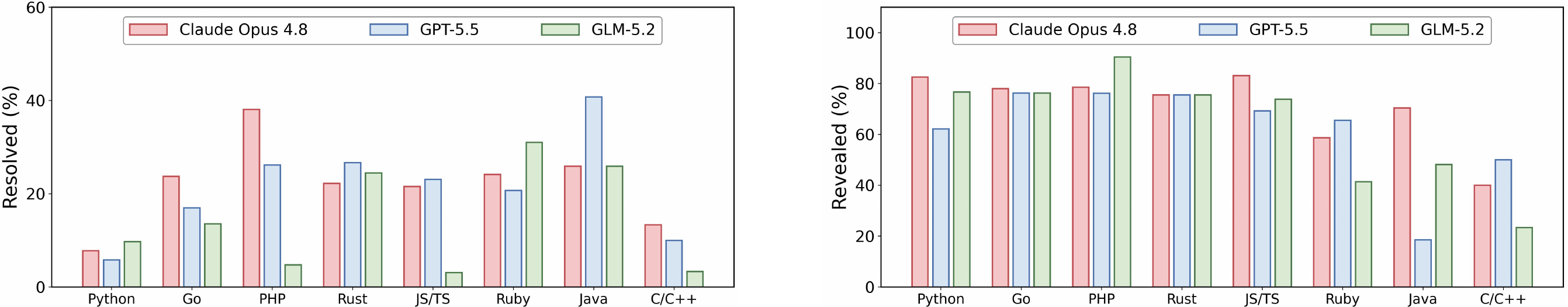}
    \caption{
    Proactive bug-fixing performance across programming languages.
    }
    \label{fig: various_language}
\end{figure*}

We have carried out experiments across various bug categories in the manuscript.
Here, we present more experimental results of proactive bug fixing across programming languages.   
As shown in Fig.~\ref{fig: various_language}, resolved rates vary substantially across programming languages, with stronger performance on PHP, Java for at least some models, while Python and C$/$C++ remain more challenging.
For potential bug discovery, the revealed rates are generally higher across most languages, but Java and C$/$C++ still show weaker results for several models.
Taken together, these conclusions suggest that recorded bug repair and potential bug discovery exhibit different language-specific patterns.

\subsection{Experimental Results under Various Scaffolds}
We conduct more experiments under different coding-agent scaffolds to examine the influence of hardness on proactive bug-fixing performance.
As shown in Table~\ref{tab: various_scaffolds}, different scaffolds lead to substantial performance variations, with Claude Code scaffold achieving better proactive bug-fixing performance.
Such results suggest that, beyond the underlying capability of LLMs, scaffold design could significantly affect proactive bug-fixing performance.

\begin{table*}[ht]
\centering
\caption{Performance comparisons under various scaffolds on a subset of Active-SWE.}
\label{tab: various_scaffolds}
\tablestyle{6pt}{1.3}
\begin{tabular}{l|cc|cc|cc}
\toprule
\multirow{2}{*}{\centering\textbf{Model}}
& \multicolumn{2}{c|}{\textbf{Claude Code}}
& \multicolumn{2}{c|}{\textbf{SWE-agent}}
& \multicolumn{2}{c}{\textbf{mini-swe-agent}} \\
& \textbf{Resolved}
& \textbf{Revealed}
& \textbf{Resolved}
& \textbf{Revealed}
& \textbf{Resolved}
& \textbf{Revealed} \\
\midrule
\modelicon{claude.png}Claude Opus 4.8 & 21.0 & 73.0 & 11.0 & 25.0 & 12.0 & 58.0 \\
\modelicon{gpt.png}GPT-5.5 & 19.0 & 63.0 & 15.0 & 32.0 & 18.0 & 46.0 \\
\modelicon{glm.png}GLM-5.2 & 12.0 & 64.0 & 10.0 & 28.0 & 11.0 & 44.0 \\
\bottomrule
\end{tabular}
\end{table*}

\subsection{Experimental Results on Cost Efficiency}

We provide additional experimental results on cost efficiency, where cache tokens are excluded for a fair comparison.
As shown in Table~\ref{tab: cost_efficiency}, proactive bug fixing exhibits a trade-off between performance and inference cost across various LLMs.
For closed-source models, Claude Opus 4.8 achieves the strongest performance but incurs the highest cost, while Qwen3.7-Max provides competitive performance at a substantially lower cost.
For open-source models, GLM-5.2 achieves the best overall performance with a relatively medium inference cost, offering a strong cost-performance trade-off.
Overall, stronger proactive bug-fixing performance often comes with higher inference cost, suggesting that effective bug discovery and repair require more extensive token usage for in-depth reasoning.

\begin{table*}[t]
\centering
\caption{Cost efficiency comparisons of different SOTA models. ``Cost'' denotes the average cost per instance.}
\label{tab: cost_efficiency}
\tablestyle{10pt}{1.3}
\begin{tabular}{l|ccc}
\toprule
\multirow{2}{*}{\centering\textbf{Model}}
& \multicolumn{3}{c}{\textbf{Cost Efficiency}} \\
& \textbf{Resolved}
& \textbf{Revealed}
& \textbf{Cost} \\
\midrule
\rowcolor{gray!12}
\multicolumn{4}{c}{\textit{\textbf{Closed-Source Models}}} \\
\midrule
\modelicon{claude.png}Claude Opus 4.8 & 20.0 & 75.0 & \$32.33 \\
\modelicon{claude.png}Claude Sonnet 4.6 & 8.3 & 53.8 & \$11.95 \\
\modelicon{gpt.png}GPT-5.5 & 18.5 & 65.5 & \$22.41 \\
\modelicon{gpt.png}GPT-5.4 & 8.8 & 32.8 & \$5.23 \\
\modelicon{gemini.png}Gemini-3.1-Pro & 15.8 & 57.0 & \$7.74 \\
\modelicon{qwen.png}Qwen3.7-Max & 11.8 & 59.8 & \$12.92 \\
\modelicon{seed.png}Seed2.1 & 11.0 & 52.8 & \$11.22 \\
\midrule
\rowcolor{gray!12}
\multicolumn{4}{c}{\textit{\textbf{Open-Source Models}}} \\
\midrule
\modelicon{deepseek.png}DeepSeek-V4-Pro & 9.5 & 53.3 & \$2.41 \\
\modelicon{glm.png}GLM-5.2 & 13.0 & 69.0 & \$8.66 \\
\modelicon{kimi.png}Kimi-K2.7-Code & 14.8 & 47.8 & \$13.54 \\
\modelicon{kimi.png}Kimi-K2.6 & 14.0 & 51.8 & \$7.53 \\
\modelicon{minimax.png}MiniMax-M2.7 & 7.8 & 38.5 & \$2.14 \\
\modelicon{ring.png}Ring-2.6-1T & 11.0 & 56.3 & \$4.18 \\
\modelicon{hy.png}Hy3 & 9.8 & 58.5 & \$0.47 \\
\modelicon{qwen.png}Qwen3.5-397B & 6.8 & 42.0 & \$1.14 \\
\modelicon{qwen.png}Qwen3.5-122B & 5.8 & 40.0 & \$0.74 \\
\modelicon{qwen.png}Qwen3.5-35B & 5.0 & 27.5 & \$0.54 \\
\bottomrule
\end{tabular}
\end{table*}

\subsection{Analytic Study on Edit Complexity}
\begin{figure*}[htbp]
    \centering
    \includegraphics[width=0.95\linewidth]{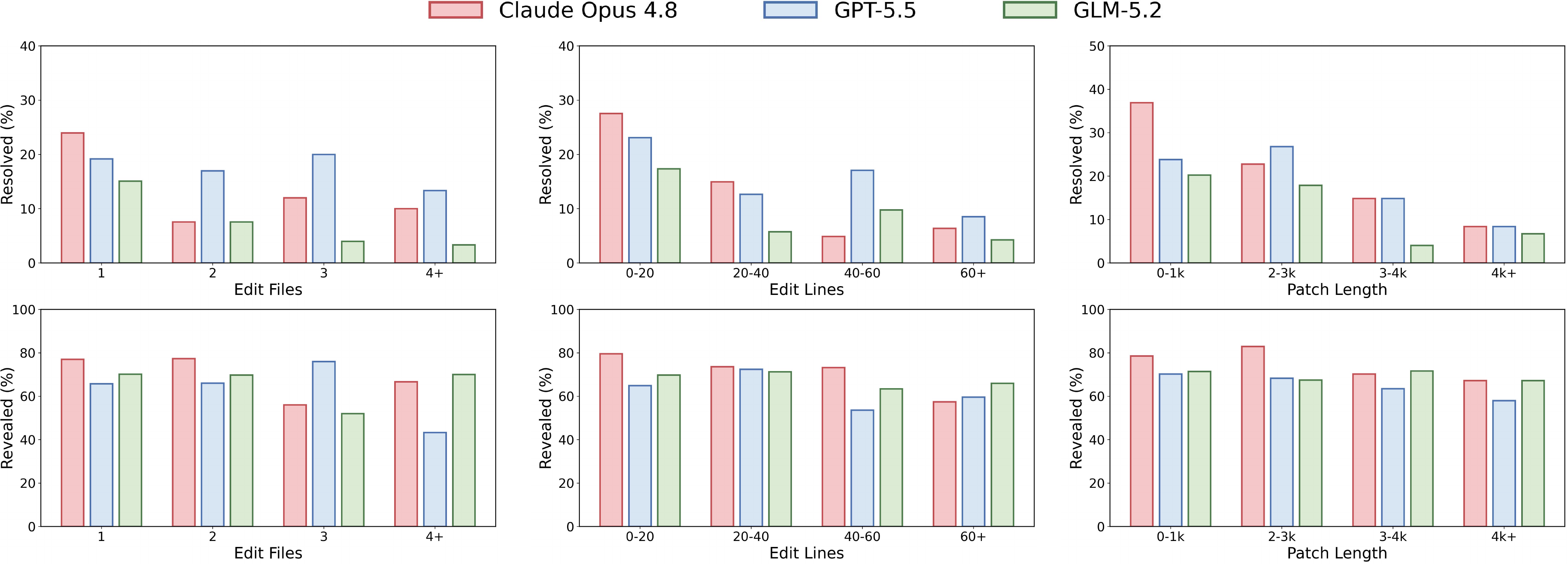}
    \caption{
    \textbf{Impact of edit complexity on proactive bug-fixing performance.}
    }
    \label{fig: edit}
\end{figure*}

We conduct additional analysis to investigate the impact of edit complexity on proactive bug fixing.
As shown in Fig.~\ref{fig: edit}, the resolved rate for recorded bugs generally decreases as the generated patch spans more edits and longer patch lengths, indicating that more modifications do not necessarily lead to correct repairs for recorded bugs.
As for potential bug discovery, the revealed rate shows a weaker and less monotonic relationship with edit complexity.

\subsection{Analytic Study on Performance across Time Periods}
\begin{figure*}[htbp]
    \centering
    \includegraphics[width=0.95\linewidth]{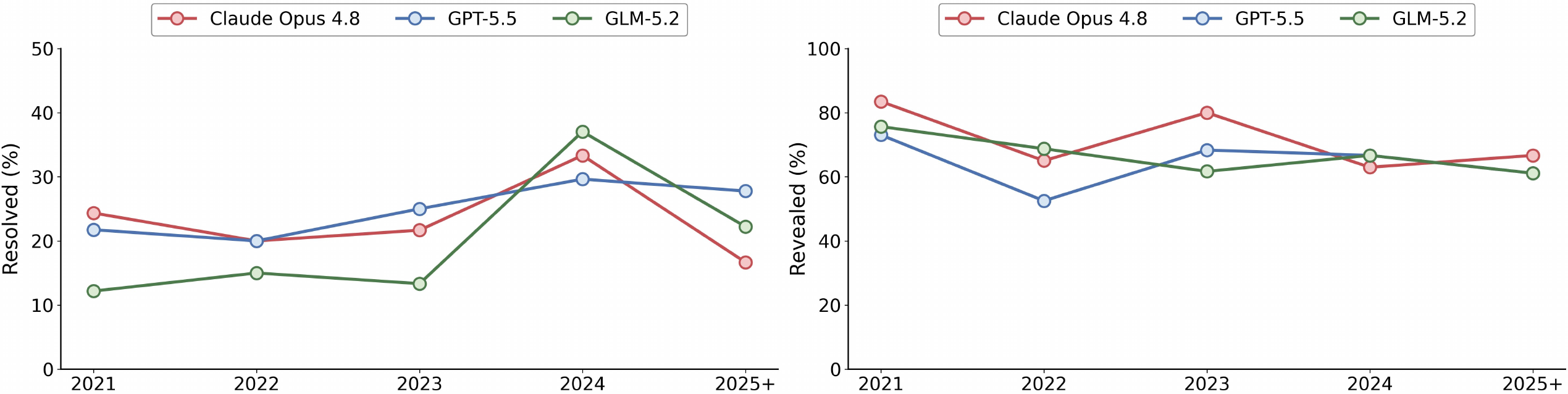}
    \caption{
    Proactive bug-fixing performance across various time periods. 
    }
    \label{fig: time}
\end{figure*}

We carry out more analysis on the performance comparison across different time periods based on the instance-wise merge time.
From the results in Fig.~\ref{fig: time}, resolved and revealed rates fluctuate across years without a consistent temporal trend, suggesting that performance is not strongly correlated with time periods.

\subsection{Analytic Study on Reactive Bug Fixing}
\begin{figure*}[t]
    \centering
    \includegraphics[width=0.9\linewidth]{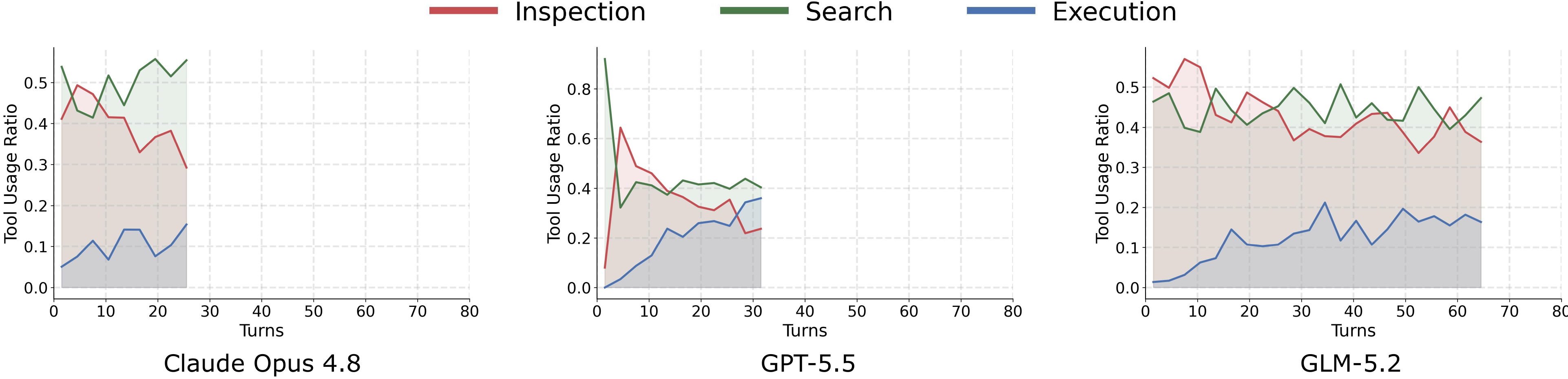}
    \caption{
    Tool-usage patterns and turn dynamics in reactive bug fixing.
    }
    \label{fig: issue_driven}
\end{figure*}

In the manuscript, we have conducted analytic study on the reasoning patterns of proactive bug fixing in Fig.~\ref{fig: issue_free}.
Here, we provide more analysis on the reasoning patterns of reactive bug fixing for comparison.
According to Fig.~\ref{fig: issue_driven}, LLMs tend to perform search in the early stages and transition faster toward execution under the reactive setting, suggesting that issue reports provide debugging cues that narrow the search space and accelerate the fixing process.
Compared with the proactive setting in Fig.~\ref{fig: issue_free}, LLMs use fewer reasoning turns when issue reports are available, indicating that additional efforts are required for in-depth exploration and analysis without issue guidance.

\subsection{Analytic Study on Broader Review Scope}
In the manuscript, we define the review scope $\mathcal{F}_i$ as the files modified by the reference patch, ensuring that $\mathcal{F}_i$ always contains the recorded bugs.
Here, we investigate how broader review scopes affect proactive bug fixing performance using the open-source SOTA model, \textit{i.e.}, GLM-5.2.
Specifically, we augment the original scope $\mathcal{F}_i$ with 2, 5, 10, or 20 additional source files, and further extend it to include all source files in the repository.
As shown in Table~\ref{tab:broader_scope}, more files pending review generally degrade performance in both recorded bug fixing and potential bug discovery, highlighting the difficulty of proactive bug fixing across broader review scopes.
In other words, even for a SOTA coding agent, proactive bug fixing becomes more challenging as the review scope grows.

\begin{table*}[ht]
    \centering
    \caption{Analytic study on broader review scopes using GLM-5.2.}
    \label{tab:broader_scope}
    \tablestyle{8pt}{1.15}
    \begin{tabular}{lcccccc}
        \toprule
        \textbf{Scope} & \textbf{LR} & \textbf{LP} & \textbf{Resolved} & \textbf{Count} & \textbf{TV} & \textbf{Revealed} \\
        \midrule
        Original         & 18.7 & 21.8 & 15.0 & 1.7 & 85.0 & 70.0 \\
        $+2$ \ Files       & 19.9 & 20.5 & 15.0 & 1.5 & 85.0 & 72.5 \\
        $+5$ \ Files       & 17.3 & 20.4 & 15.0 & 1.9 & 85.0 & 65.0 \\
        $+10$ \ Files      & 16.5 & 13.4 & 12.5 & 1.5 & 77.5 & 65.0 \\
        $+20$ \ Files     & 9.8  & 7.2  & 10.0 & 1.8 & 80.0 & 60.0 \\
        All Files & 4.0  & 1.4  & 2.5  & 3.7 & 47.5 & 15.0 \\
        \bottomrule
    \end{tabular}
\end{table*}

\subsection{Analytic Study on Different Judge Agents}
To assess the sensitivity of potential bug evaluation to judge selection, we compare the evaluation results of two open-source SOTA models, Qwen3.5-397B and GLM-5.2, with those of human experts.
From the results in Table~\ref{tab:various_judges}, the judge agents driven by various LLMs and human experts yield similar evaluation results, suggesting limited sensitivity to the selection of judge agents.

\begin{table*}[htbp]
    \centering
    \caption{Potential bug evaluation with different judges.}
    \label{tab:various_judges}
    \tablestyle{8pt}{1.15}
    \begin{tabular}{l|cc}
        \toprule
        \textbf{Judge}
        & \textbf{Count}
        & \textbf{Revealed} \\
        \midrule
        Qwen3.5-397B & 1.7 & 69.0 \\
        GLM-5.2 & 1.7 & 70.0 \\
        Human Expert & 1.7 & 68.0 \\
        \bottomrule
    \end{tabular}
\end{table*}

\subsection{Analytic Study on the Overlap Between Potential and Recorded Bugs}
To investigate whether revealed potential bugs extend beyond the recorded set, we ask human experts to examine the overlap between potential and recorded bugs.
As demonstrated in Table~\ref{tab:potential_recorded_overlap}, for all the coding agents, the revealed bugs consistently go beyond the recorded ones, suggesting that coding agents are able to discover potentially unrecorded bugs.

\begin{table*}[ht]
    \centering
    \caption{Comparison between the validated potential bugs and the recorded bug. ``Coverage'' denotes whether at least one potential bug is validated. ``Overlap'' indicates the ratio that valid potential bugs matching recorded bugs.
    }
    \label{tab:potential_recorded_overlap}
    \tablestyle{8pt}{1.15}
    \begin{tabular}{l|ccc}
        \toprule
        \textbf{Model}
        & \textbf{Valid Bugs}
        & \textbf{Coverage}
        & \textbf{Overlap} \\
        \midrule
        GLM-5.2 & 1.7 & 90.0 & 15.6 \\
        Claude Opus 4.8 & 1.2 & 85.0 & 48.0 \\
        GPT-5.5 & 3.0 & 90.0 & 9.6 \\
        \bottomrule
    \end{tabular}
\end{table*}

\newpage

\section{Task Templates}
In this section, we present more details on the task templates used throughout Active-SWE construction and evaluation, including the taxonomy curation template $\mathcal{T}_{\text{C}}$, the bug fixing template $\mathcal{T}_{\text{R}}$, the test generation template $\mathcal{T}_{\text{S}}$, and the judge template $\mathcal{T}_{\text{J}}$.

\begin{tcolorbox}[
    title=Taxonomy Curation Template,
    colback=gray!5,
    colframe=black!60,
    fonttitle=\bfseries,
    breakable
]

\textbf{Role}

You are a senior software engineer helping curate a benchmark for bug fixing.

\vspace{1.0em}

\textbf{Objective}

Given a PR with its issue report, code patch, and test patch, your task is to decide whether it fixes a valid bug that falls into the target taxonomy.
Exclude PRs whose root causes fall outside the taxonomy, such as feature requests, documentation changes, refactoring, test-only fixes, and performance optimizations without correctness impact.
If the PR fixes a target bug, classify the bug into exactly one category from the target taxonomy.

\vspace{1.0em}

\textbf{Input}

\begin{itemize}
    \item \textbf{[Issue Report]}
    \item \textbf{[Code Patch]}
    \item \textbf{[Test Patch]}
\end{itemize}

\vspace{1.0em}

\textbf{Bug Taxonomy}

\textbf{[Bug Taxonomy]}

\end{tcolorbox}

\newpage

\begin{tcolorbox}[
    title=Bug Taxonomy,
    colback=gray!5,
    colframe=black!60,
    fonttitle=\bfseries,
    breakable
]

\textbf{Logic \& Computation}
\begin{itemize}
    \item \textbf{Incorrect Condition / Inverted Logic}: incorrect comparison, boolean operator, negation, or branch condition in an existing conditional.
    \item \textbf{Incomplete Conditional}: missing a necessary branch, case, or variant within an existing \texttt{if}/\texttt{switch} structure.
    \item \textbf{Expression / Computation Error}: arithmetic, precision, operator-precedence, overflow, boundary, or off-by-one errors.
\end{itemize}

\textbf{Reference \& Data Flow}
\begin{itemize}
    \item \textbf{Incorrect Variable / Property / Key}: use of an incorrect variable, field, property, attribute, or dictionary key.
    \item \textbf{Incorrect Method / API Call}: invocation of an incorrect API variant, constructor, operator, or similarly named method.
    \item \textbf{Missing Propagation / Context Loss}: failure to propagate required parameters, configuration, metadata, state, or context across calls.
\end{itemize}

\textbf{Domain-specific Workflow}
\begin{itemize}
    \item \textbf{Missing Business-Rule Guard}: absence of a required domain constraint, policy check, permission check, or feature guard.
    \item \textbf{Incomplete Registry / Allowlist}: missing a valid entry in an existing registry, mapping, allowlist, or configuration table.
    \item \textbf{Incorrect Operation Ordering}: operations executed in an invalid, priority-violating, or rule-violating order.
\end{itemize}

\textbf{Data Processing \& Type}
\begin{itemize}
    \item \textbf{String / Regex / Path Error}: incorrect string manipulation, regular-expression matching, delimiter handling, escaping, or path/URL construction.
    \item \textbf{Missing Parser / Grammar Rule}: missing token, grammar rule, syntax definition, or parser/compiler handling.
    \item \textbf{Type Mismatch / Implicit Conversion}: incorrect type assumptions, unintended implicit conversions, or missing type-specific handling.
\end{itemize}

\textbf{Exception Safety}
\begin{itemize}
    \item \textbf{Missing Null / Existence Check}: missing checks for \texttt{null}/\texttt{None}/\texttt{nil}/\texttt{undefined}, absent keys, empty collections, or valid falsy values.
    \item \textbf{Unhandled Exception}: failure to catch, handle, recover from, or propagate an expected exception, panic, rejection, or error path.
    \item \textbf{Silent Exception}: exception handling that masks, swallows, fails to log, or fails to propagate an error.
\end{itemize}

\textbf{State \& Lifecycle}
\begin{itemize}
    \item \textbf{Incorrect State Transition}: incorrect state update, invalid transition, or violation of readiness/order dependencies.
    \item \textbf{Stale Data / Concurrency Safety}: stale cached or closure-captured data, race conditions, unsynchronized shared state, or deadlock-prone concurrency.
    \item \textbf{Resource Lifecycle Error}: improper acquisition, release, cancellation, cleanup, or lifetime management of resources.
\end{itemize}

\end{tcolorbox}

\newpage

\begin{tcolorbox}[
    title=Proactive Bug Fixing Template,
    colback=gray!5,
    colframe=black!60,
    fonttitle=\bfseries,
    breakable
]

\textbf{Role}

You are a senior software engineer performing proactive bug fixing.

\vspace{1.0em}

\textbf{Objective}

Given the list of \textbf{files pending review}, your task is to identify the high-impact bugs and provide the corresponding fixes. 
Ignore all non-functional concerns, including style, formatting, and naming, to focus solely on functional correctness and system integrity.


\vspace{1.0em}

\textbf{Review Dimensions}

The following are the primary bug categories to be reviewed:

\textbf{[Target Bug Taxonomy]}

\vspace{1.0em}

\textbf{Multi-Bug Review Protocol}

The provided files pending review in the repo may contain multiple independent critical failures. 
You must execute an exhaustive code review:

\begin{itemize}
    \item \textbf{No Early Exit}: Scan exhaustively. Do not stop after finding the first bug.
    \item \textbf{Surgical Isolation}: Document and fix each bug independently. Never combine unrelated patches into a monolithic refactor.
\end{itemize}

\vspace{1.0em}

\textbf{Input}

\begin{itemize}
    \item \textbf{[Files Pending Review]}
\end{itemize}

\end{tcolorbox}

\begin{tcolorbox}[
    title=Test Generation Template,
    colback=gray!5,
    colframe=black!60,
    fonttitle=\bfseries,
    breakable
]

\textbf{Role}

You are a senior software engineer generating tests to reproduce revealed bugs.

\vspace{1.0em}

\textbf{Objective}

You have developed code patch that may fix one or more independent bugs during the bug-fixing task.
Now, your task is to write exactly one test function for each bug fixed in the code patch.
Each test function must satisfy the following requirements:

\begin{itemize}
    \item Fails on the buggy codebase before the patch is applied. 
    \item Passes on the fixed codebase after the patch is applied.
\end{itemize}

Do NOT write tests that verify normal or already-correct behavior.
Every generated test must expose a bug fixed by the code patch.

\vspace{1.0em}

\textbf{Input}

\begin{itemize}
    \item \textbf{[Proactive Bug Fixing Template]}
    \item \textbf{[Predicted Code Patch]}
\end{itemize}

\end{tcolorbox}

\newpage

\begin{tcolorbox}[
    title=Judge Task Template,
    colback=gray!5,
    colframe=black!60,
    fonttitle=\bfseries,
    breakable
]

\textbf{Role}

You are a senior software engineer for validating revealed bugs and generated tests.

\vspace{1.0em}

\textbf{Objective}

Your task is to evaluate whether the predicted code patch fixes valid bugs and whether the provided test patch truly reproduces these bugs. 

\vspace{1.0em}

\textbf{Key Definitions}

\begin{itemize}
    \item \textbf{Effective Bug}: a real functional bug that is fixed by the predicted code patch. Cosmetic edits, formatting changes, pure refactoring, dead-code changes, test-only changes, and undocumented behavior changes are NOT effective bug fixes.
    \item \textbf{Effective Test}: a test that reproduces an effective bug. Tests are NOT effective if they only assert patched outputs, duplicate implementation logic, or check behavior unrelated to the bug.
\end{itemize}

\vspace{1.0em}

\textbf{Bug Taxonomy Mapping}

Map each bug in the predicted code patch to one of the 7 taxonomies:
\begin{enumerate}
    \item Logic \& Computation
    \item Reference \& Data Flow
    \item Domain-specific Workflow
    \item Data Processing \& Type
    \item Exception Safety
    \item State \& Lifecycle
    \item Others
\end{enumerate}

\vspace{1.0em}

\textbf{Evaluation Protocol}

You must execute a rigorous evaluation based on the following rules:
\begin{itemize}
    \item \textbf{Bug Identification}: identify each distinct bug fixed in the predicted patch. Multiple fixes addressing the same root cause must be grouped into ONE effective bug.
    \item \textbf{Taxonomy Mapping}: map each identified effective bug to exactly one bug taxonomy.
    \item \textbf{Test Validation}: determine the effectiveness of each Fail-to-Pass (F2P) test and explicitly link it to the specific bug it verifies.
\end{itemize}

\vspace{1.0em}

\textbf{Input}

\begin{itemize}
    \item \textbf{[Proactive Bug Fixing Template]}
    \item \textbf{[Predicted Code Patch]}
    \item \textbf{[Predicted Test Patch]}
    \item \textbf{[Fail-to-Pass Tests]}

\end{itemize}

\end{tcolorbox}

\newpage
\section{Case Study}
\label{sec:case-study}

In this section, we conduct case studies on proactive bug-fixing tasks.
As shown in Fig.~\ref{fig:case-recorded-success}-\ref{fig:case-potential-incomplete}, we provide successful and failed cases in recorded bug fixing and potential bug discovery.

\begin{center}
    \centering
    \includegraphics[width=0.65\linewidth]{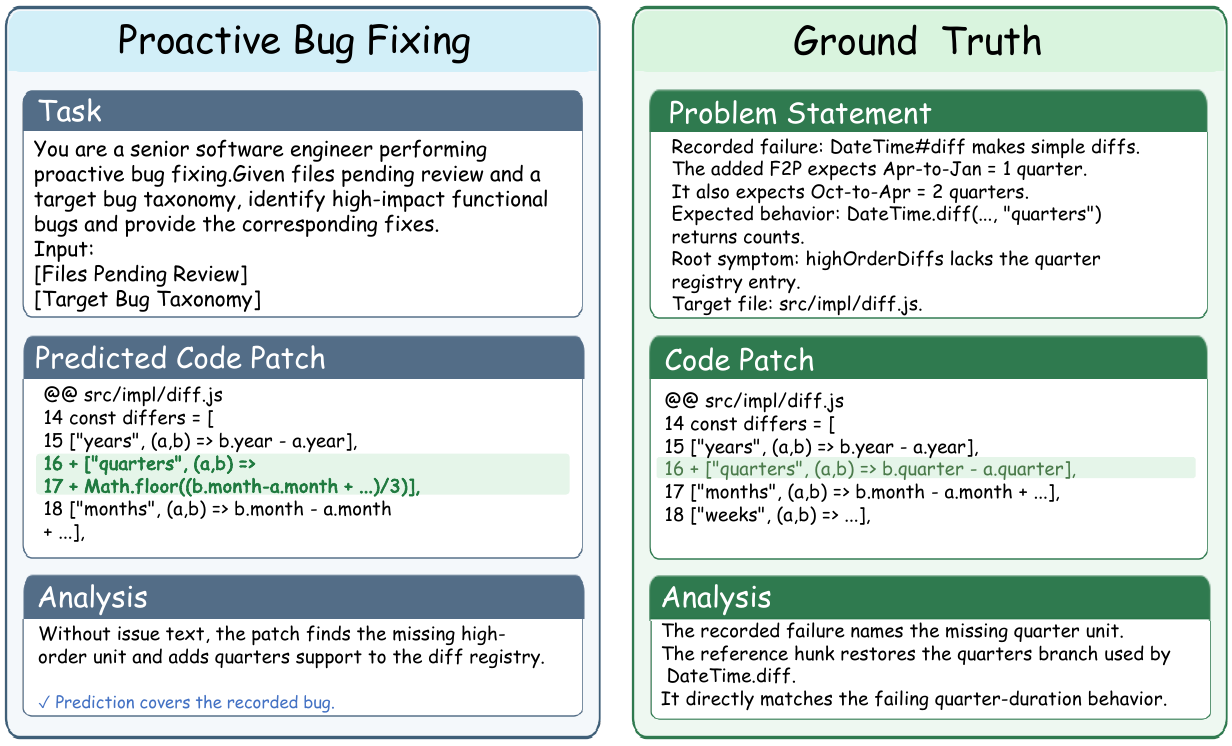}
    \captionsetup{hypcap=false}
    \captionof{figure}{Successful recorded-bug fixing case where the predicted patch fixes the missing
    quarters branch.}
    \label{fig:case-recorded-success}
\end{center}

\begin{center}
    \centering
    \includegraphics[width=0.65\linewidth]{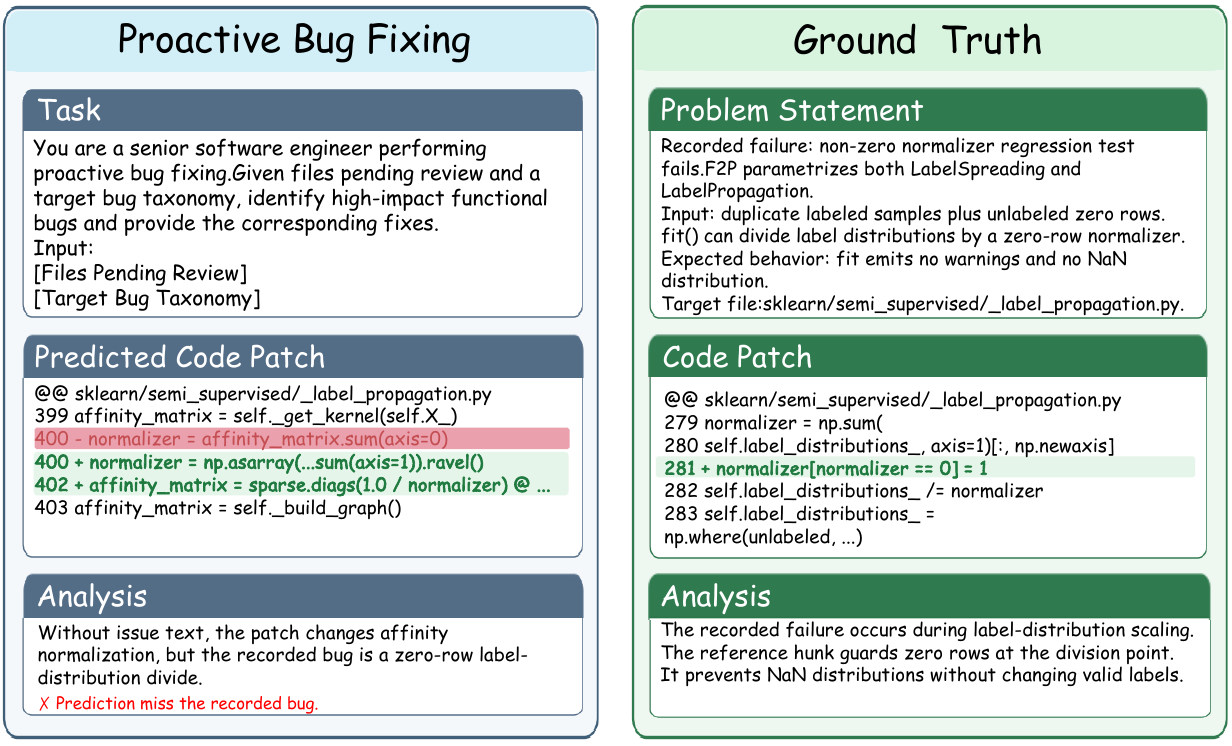}
    \captionsetup{hypcap=false}
    \captionof{figure}{Failed recorded-bug fixing case where the predicted patch misses the label-distribution
    scaling failure.}
    \label{fig:case-recorded-failure}
\end{center}

\begin{center}
    \centering
    \includegraphics[width=\linewidth]{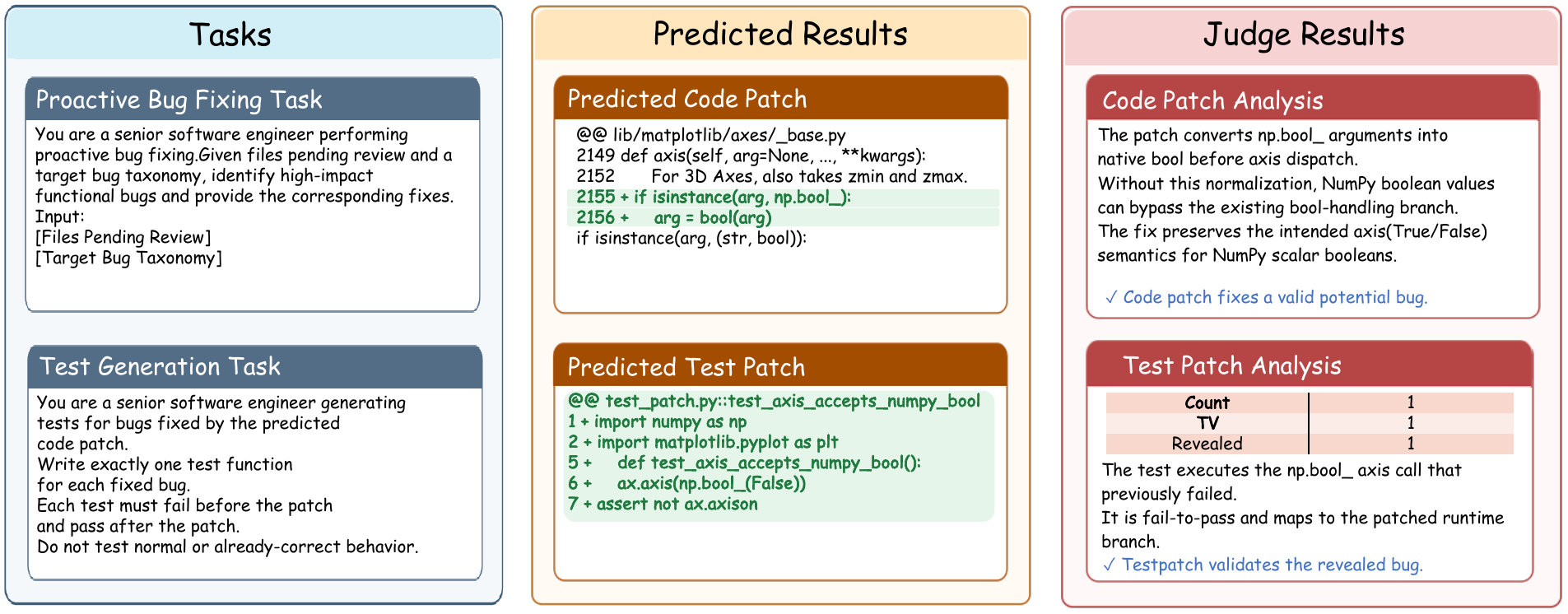}
    \captionsetup{hypcap=false}
    \captionof{figure}{Successful potential bug discovery case where the generated tests validate the revealed
    bug.}
    \label{fig:case-potential-success}
\end{center}

\begin{center}
    \centering
    \includegraphics[width=\linewidth]{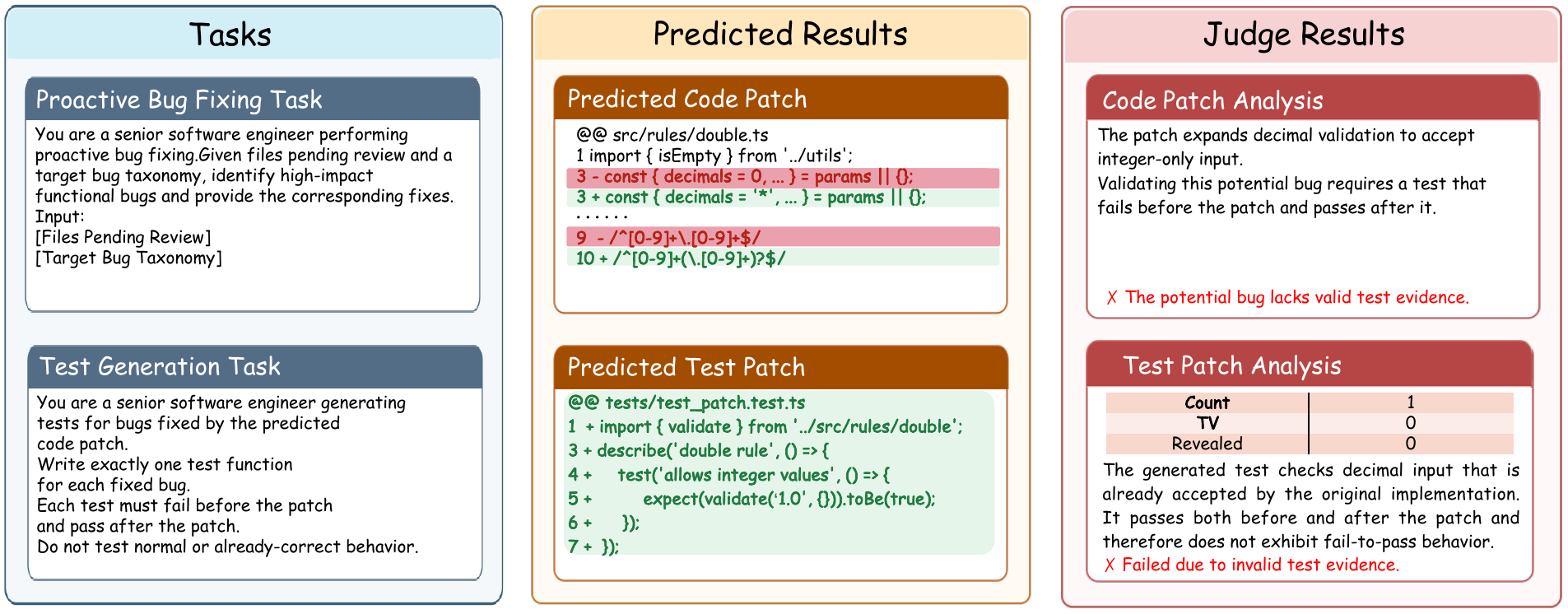}
    \captionsetup{hypcap=false}
    \captionof{figure}{Failed potential bug discovery case where the generated tests don't exhibit fail-to-pass behavior.}
    \label{fig:case-potential-invalid-claim}
\end{center}

\begin{center}
    \centering
    \includegraphics[width=\linewidth]{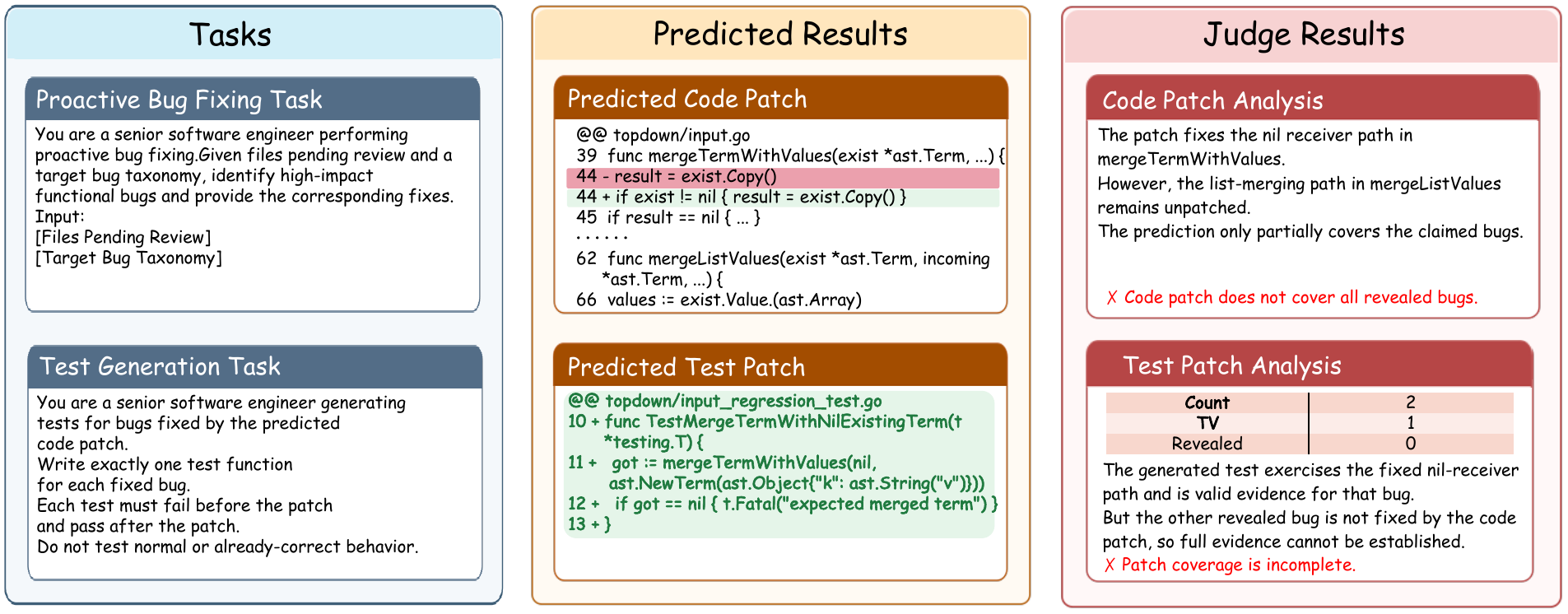}
    \captionsetup{hypcap=false}
    \captionof{figure}{Failed potential bug discovery case 
    where the generated tests don't cover all the revealed bugs.}
    \label{fig:case-potential-incomplete}
\end{center}

\end{document}